\documentclass[twocolumn, superscriptaddress, preprintnumbers,amsmath,amssymb]{revtex4}

\usepackage{booktabs}
\usepackage{color,graphicx}
\usepackage{amsmath}
\usepackage{dcolumn}
\usepackage{CJK}                 
\usepackage{bm}                  
\usepackage{soul}
\usepackage{enumerate}
\usepackage{multirow}
\usepackage{mathrsfs}
\usepackage{braket}
\usepackage{longtable}
\usepackage{pdflscape}

\usepackage[dvipdfm,
            pdfstartview=Fit,
            CJKbookmarks=true,
            bookmarksnumbered=true,
            bookmarksopen=true,
            colorlinks,
            pdfborder=001,
            linkcolor=blue,
            anchorcolor=blue,
            citecolor=blue
            ]{hyperref}

\usepackage{float}

\begin{document}
\begin{CJK*}{GBK}{song}

\title{Possible neutron halo in triaxial nucleus $^{42}$Al}

\author{K. Y. Zhang}
\affiliation{State Key Laboratory of Nuclear Physics and Technology, School of Physics, Peking University, Beijing 100871, China}
\affiliation{Institute of Nuclear Physics and Chemistry, CAEP, Mianyang, Sichuan 621900, China}

\author{S. Q. Zhang}
\affiliation{State Key Laboratory of Nuclear Physics and Technology, School of Physics, Peking University, Beijing 100871, China}

\author{J. Meng} \email{mengj@pku.edu.cn}
\affiliation{State Key Laboratory of Nuclear Physics and Technology, School of Physics, Peking University, Beijing 100871, China}

\begin{abstract}
  A  microscopic self-consistent triaxial relativistic Hartree-Bogoliubov theory in continuum (TRHBc), which simultaneously takes into account the triaxiality and pairing correlations as well as continuum effects, is established and applied to explore the novel halo phenomenon in aluminum isotopes.
  The experimental proton drip line and the available data of neutron separation energies and charge radii are reproduced well without any free parameters.
  The neutron-richest odd-odd aluminum isotope observed so far, $^{42}$Al, is predicted to be triaxially deformed with $\beta=0.35$ and $\gamma=42^\circ$.
  Its one-neutron separation energy is predicted to be $0.68$ MeV, in agreement with the AME2020, and the neutron rms radius is 3.94 fm, remarkably larger than the empirical value.
  The density distribution of the valance neutron, which extends much farther in space than the core, suggests a possible neutron halo in $^{42}$Al.
  The dominant components responsible for the spatial extension of the halo are revealed by the single-neutron orbitals around the Fermi energy.
  A novel phenomenon, the exchange of the intermediate and short axes between the triaxial core with $\beta=0.38$ and $\gamma=50^\circ$, and the triaxial halo with $\beta=0.79$ and $\gamma=-23^\circ$, is found.
  Future experiments to explore the halo phenomenon and the novel shape decoupling in $^{42}$Al are highly demanded.
\end{abstract}

\date{\today}

\maketitle


Quantum halo systems, characterized by substantial components extending well into classically forbidden regions,
are of particular interests in molecular, atomic, and nuclear physics~\cite{Jensen2004RMP}.
The halo in nuclear physics starts from the interaction cross section measurement of Li isotopes on target $^{12}$C~\cite{Tanihata1985PRL}, which become the driving force behind the worldwide radioactive ion beam facilities.

The discovery of the halo in nuclear physics provides a challenge to the conventional theory of nuclear structure, because the weakly bound nuclei involve the coupling between bound states and the continuum.
A fully microscopic and self-consistent explanation of the neutron halo in $^{11}$Li is provided by the relativistic continuum Hartree-Bogoliubov (RCHB) theory~\cite{Meng1996PRL}, which couples bound states and the continuum by pairing correlations.
A novel phenomenon, \emph{giant halo}, formed by up to six neutrons, has also been predicted by the RCHB theory in Zr isotopes near the neutron drip line~\cite{Meng1998PRL}.

The existence of halo in deformed nuclei had been under debate for decades~\cite{Tanihata1995NPA,Hamamoto2004PRC(R),Nunes2005NPA,Zhou2010PRC(R)}.
Based on a spherical Woods-Saxon potential, the drip-line nuclei are suggested to be spherical~\cite{Tanihata1995NPA}.
Based on an axially deformed Woods-Saxon potential, the existence of deformed halos is doubted because the $s$ wave component becomes dominant in the wave functions of $\Omega^\pi = 1/2^+$ orbitals as their binding energies approach zero~\cite{Hamamoto2004PRC(R)}.
Based on a three-body model, the formation of a deformed halo near the drip line is suggested to be unlikely~\cite{Nunes2005NPA}.

In 2010, a microscopic deformed relativistic Hartree-Bogoliubov theory in continuum (DRHBc) was developed, which self-consistently takes into account the axial deformation, pairing correlations, and continuum effects~\cite{Zhou2010PRC(R)}.
The deformed halos in neutron-rich Mg isotopes are predicted and shape decoupling between the core and the halo is illustrated by the DRHBc theory~\cite{Zhou2010PRC(R),Li2012PRC}.
In 2014, the experimental evidence for deformed halos was reported in $^{31}$Ne~\cite{Nakamura2014PRL} and $^{37}$Mg~\cite{Kobayashi2014PRL}.
Recently, the DRHBc theory has been applied to investigate halo phenomena in
$^{17,19}$B~\cite{Yang2021PRL,Sun2021PRC(1)},
$^{15,19,22}$C~\cite{Sun2018PLB,Sun2020NPA},
$^{31}$Ne~\cite{Zhong2022SciChina},
and $^{42,44}$Mg~\cite{Zhou2010PRC(R),Li2012PRC,Zhang2019PRC,Sun2021SciBull}.

The existence of the halo phenomenon in triaxial nuclei is an interesting but less explored topic.
In particular, the importance of triaxiality has been demonstrated in nuclear fission~\cite{Lyu2012PRC(R)} and novel phenomena such as the nuclear chirality~\cite{Frauendorf1997NPA} and wobbling motion~\cite{Bohr1975Book}.
Recently, based on a Woods-Saxon potential, it is pointed out that the region of halo nuclei might be extended because the triaxial deformation allows the appearance of $s$ or $p$ wave components in some weakly bound orbitals~\cite{Uzawa2021PRC(L)}.
Further investigation is definitely crucial to include pairing correlations and continuum effects as well as self-consistent triaxiality.

In this Letter, a microscopic triaxial relativistic Hartree-Bogoliubov theory in continuum (TRHBc), which includes self-consistently the triaxiality, pairing correlations, and continuum effects, is developed and applied to explore the halo phenomenon in triaxial nuclei.


In the TRHBc theory, the relativistic Hartree-Bogoliubov equations for the nucleons read~\cite{Kucharek1991ZPA}
\begin{equation}\label{RHB}
      \left(\begin{matrix}
          h_D-\lambda   &    \Delta \\
          -\Delta^*         &   -h_D^*+\lambda
      \end{matrix}\right)
      \left(\begin{matrix}
          U_k\\
          V_k
      \end{matrix}\right)
  = E_k
     \left(\begin{matrix}
      U_k\\
      V_k
     \end{matrix}\right),
\end{equation}
in which $\lambda$ is the Fermi energy, and $E_k$ and $(U_k, V_k)^{\rm T}$ are the quasiparticle energy and wave function, respectively.
The quasiparticle wave function is expanded in a Dirac Woods-Saxon basis~\cite{Zhou2003PRC,Zhang2022PRC}, which can describe the large spatial extension of halo nuclei.
$h_D$ is the Dirac Hamiltonian,
\begin{equation}
      h_D(\bm{r})=\bm{\alpha}\cdot\bm{p}+V(\bm{r})+\beta[M+S(\bm{r})],
\end{equation}
with the scalar potential $S(\bm r)$ and vector potential $V(\bm r)$ constructed from the quasiparticle wave functions.
The pairing potential (neglecting the spin and isospin indexes for simplicity) reads
\begin{equation}\label{Delta}
      \Delta(\bm r_1,\bm r_2) = V^{\mathrm{pp}}(\bm r_1,\bm r_2)\kappa(\bm r_1,\bm r_2),
\end{equation}
with a density-dependent force of zero range,
\begin{equation}\label{pair}
      V^{\mathrm{pp}}(\bm r_1,\bm r_2)
  = V_0 \frac{1}{2}(1-P^\sigma)\delta(\bm r_1-\bm r_2)\left(1-\frac{\rho(\bm r_1)}{\rho_{\mathrm{sat}}}\right),
\end{equation}
and the pairing tensor $\kappa$~\cite{Peter1980Book}.

The potentials and densities are expanded in terms of spherical harmonic functions,
\begin{equation}\label{legendre}
      f(\bm r)  =  \sum_{\lambda\mu} f_{\lambda\mu}(r)Y_{\lambda\mu}(\theta,\varphi),
\end{equation}
where $\lambda=0,2,4,\cdots$ and $\mu = -\lambda,-\lambda+2,\cdots,\lambda$ are restricted by spatial reflection symmetry and simplex symmetry.


Equation (\ref{RHB}) is solved self-consistently for aluminum isotopes with density functionals PC-PK1~\cite{Zhao2010PRC}, NL3*~\cite{Lalazissis2009PLB}, NLSH~\cite{Sharma1993PLB}, and PK1~\cite{Long2004PRC}.
In Eq.~(\ref{pair}), the pairing strength $V_0=-342.5~\mathrm{MeV~fm}^3$ and the saturation density $\rho_{\mathrm{sat}}=0.152~\mathrm{fm}^{-3}$, and a pairing window of $100$ MeV is adopted, the same as those in the global RCHB calculations over the nuclear chart~\cite{Xia2018ADNDT}.
For the Dirac Woods-Saxon basis, the energy cutoff $E^+_{\mathrm{cut}}=300$ MeV and the angular momentum cutoff $J_{\max}=19/2~\hbar$ are adopted, which have been proved to provide converged results~\cite{Zhang2020PRC}.
In Eq.~(\ref{legendre}), the spherical harmonic expansion truncation is chosen as $\lambda_{\max}=6$~\cite{Zhang2020PRC,Pan2019IJMPE,Zhang2022ADNDT}.
The blocking effects of odd nucleon(s) are taken into account via the equal filling approximation~\cite{Perez-Martin2008PRC,Li2012CPL,Pan2022PRC}.


\begin{figure}[htbp]
  \centering
  \includegraphics[width=0.4\textwidth]{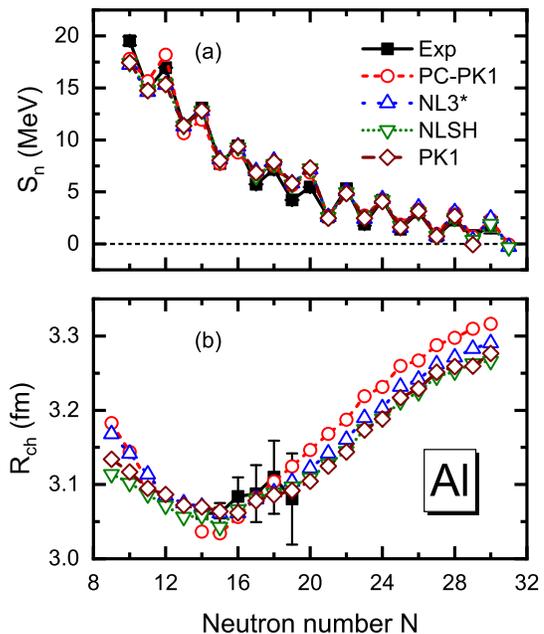}
  \caption{(a) One-neutron separation energy $S_n$ and (b) charge radius $R_\mathrm{ch}$ as functions of the neutron number $N$
                 for aluminum isotopes from TRHBc calculations, in comparison with data available~\cite{AME2020(3),Heylen2021PRC}.}
\label{fig1}
\end{figure}

In Fig.~\ref{fig1}, the one-neutron separation energy $S_n$ and charge radius $R_\mathrm{ch}$ calculated by the TRHBc theory with PC-PK1 for aluminum isotopes are given from the proton drip line to the one-neutron drip line, in comparison with data available~\cite{AME2020(3),Heylen2021PRC} and results with NL3*, NLSH, and PK1.

In Fig.~\ref{fig1}(a), the one-neutron separation energy $S_n$ from AME2020~\cite{AME2020(3)} and its odd-even staggering are reproduced both in tendency and magnitude.
In Fig.~\ref{fig1}(b), the recently measured charge radii~\cite{Heylen2021PRC} are well reproduced by the TRHBc theory within the experimental uncertainty.
Since the high density instability~\cite{Reinhard1988ZPA,Zhang2022ADNDT} occurs for PC-PK1 near the neutron number $N=12$, the results for $^{25,26}$Al are not shown.
On the proton-rich side, the experimental proton drip-line nucleus $^{22}$Al~\cite{AME2020(3)} is correctly reproduced by the four density functionals.
On the neutron-rich side,  the one-neutron drip-line nucleus is predicted as $^{43}$Al by PC-PK1, NL3*, and NLSH, while as $^{41}$Al by PK1, which is to be confirmed in future experiment.
$^{42}$Al is the neutron-richest odd-odd aluminum isotope observed so far~\cite{Baumann2007Nature}.
Its empirical one-neutron separation energy, $S_n$ = $0.67(64)$ MeV~\cite{AME2020(3)}, nicely reproduced by the TRHBc theory, might be a signal of one-neutron halo.

\begin{figure*}[htbp]
  \centering
  \includegraphics[width=0.7\textwidth]{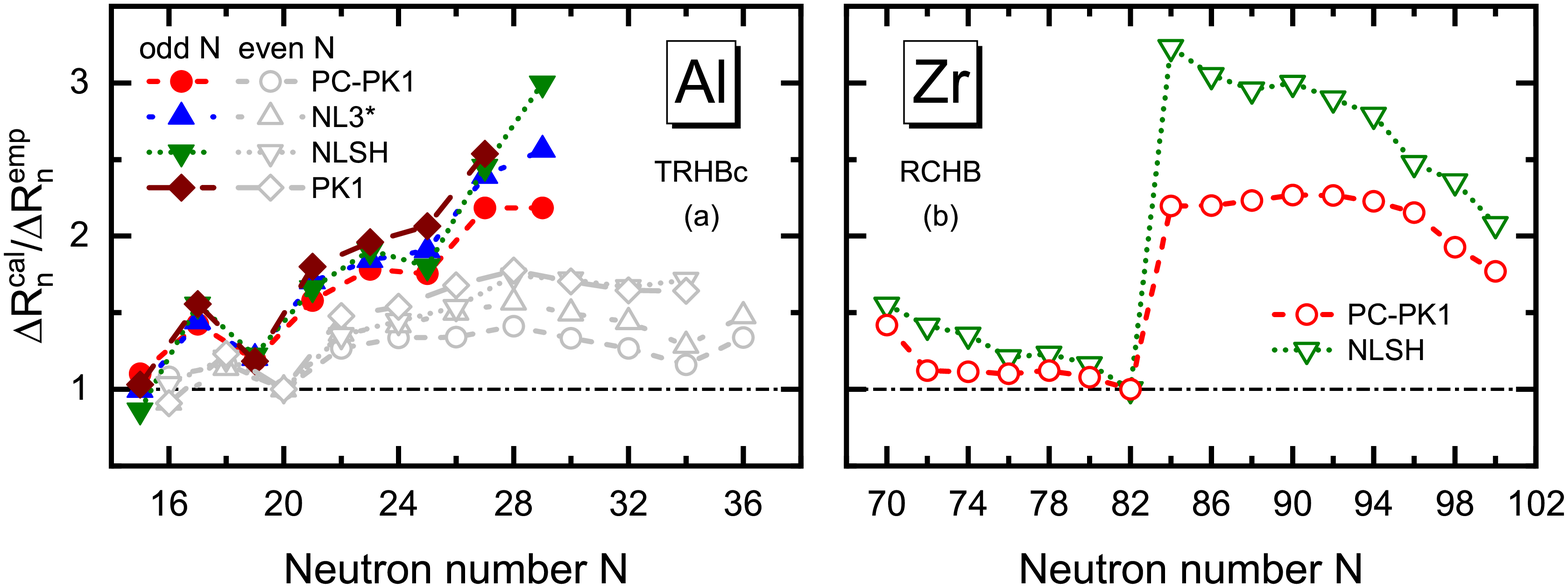}
  \caption{The halo scale $S_{\mathrm{halo}}=\Delta R_n^{\mathrm{cal}}/\Delta R_n^{\mathrm{emp}}$ as a function of the neutron number
                 $N$ for (a) aluminum isotopes in TRHBc calculations and (b) zirconium isotopes in RCHB calculations~\cite{Meng1998PRL,Xia2018ADNDT}. For the empirical values, $r_0$ is determined by renormalizing $S_{\mathrm{halo}}$ as one at $N=20$ in (a) and at $N=82$ in (b).}
\label{fig2}
\end{figure*}

For the neutron halo in $^{11}$Li, the two valance neutrons increase the matter radius from less than 2.4 fm of $^{9}$Li to around 3.5 fm of $^{11}$Li~\cite{Tanihata2013PPNP}.
For medium and heavy nuclei, however, the impact of one or two halo neutrons would be less prominent.
The characterization of the halo phenomenon in medium and heavy nuclei remains a hot topic in nuclear physics for past decades~\cite{Meng2015JPG}.

Here we propose a new \emph{halo scale} to characterize the halo phenomenon in medium and heavy nuclei.
The main idea is to compare the contribution to the rms radius by the weakly bound neutron(s) with the conventional one.
Empirically, the neutron rms radius of a nucleus $R_n^{\mathrm{emp}} (N) = r_0N^{1/3}$.
Adding $m$ neutrons, the increment of the neutron rms radius is $\Delta R_n^{\mathrm{emp}} = R_n^{\mathrm{emp}} (N + m ) - R_n^{\mathrm{emp}} (N)$.
From the experimental neutron rms radius $R_n^{\mathrm{exp}}$ or that in microscopic calculations $R_n^{\mathrm{cal}}$, the halo scale is defined as
\begin{equation}\label{HS}
 S_{\mathrm{halo}}= \frac{\Delta R_n^{\mathrm{exp(cal)}}}{\Delta R_n^{\mathrm{emp}}}= \frac{R_n^{\mathrm{exp(cal)}} (N + m ) - R_n^{\mathrm{exp(cal)}} (N)}{R_n^{\mathrm{emp}} (N + m ) - R_n^{\mathrm{emp}} (N)},
\end{equation}
where $m = 1$ can be used for the one-neutron halo, $m = 2$ the two-neutron halo, and so on.
An enhancement of  $S_{\mathrm{halo}}$ might be regarded as a signal of the halo phenomenon.

The halo scale $S_{\mathrm{halo}}$ for neutron-rich aluminum isotopes by the TRHBc theory is shown in Fig.~\ref{fig2}(a), in comparison with the results for neutron-rich zirconium isotopes by the RCHB theory~\cite{Meng1998PRL,Xia2018ADNDT} in Fig.~\ref{fig2}(b).
As shown in Fig.~\ref{fig2}(b), a sudden increase and large values of $S_{\mathrm{halo}}$ after $N=82$ are found in the predicted region of halo nuclei~\cite{Meng1998PRL}.
This demonstrates the validity of the defined halo scale as a signal for halo nuclei.
In Fig.~\ref{fig2}(a), the sudden increase and the magnitude of $S_{\mathrm{halo}}$ for $^{40}$Al and $^{42}$Al are comparable with those for halo nuclei in zirconium isotopes.
With a smaller $S_n$ and a relatively large $S_{\mathrm{halo}}$, the nucleus $^{42}$Al will be investigated in detail for possible halo structure.

\begin{figure}[htbp]
  \centering
  \includegraphics[width=0.35\textwidth]{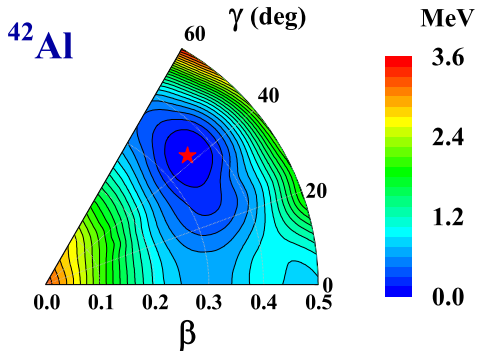}
  \caption{Potential energy surface for $^{42}$Al in the $\beta$-$\gamma$ plane from constrained TRHBc calculations with density functional
                 PC-PK1. All energies are normalized with respect to the energy of the absolute minimum (in MeV) indicated by the star. The energy separation between each contour line is 0.15 MeV.}
\label{fig3}
\end{figure}

The possible triaxial shape of $^{42}$Al can be revealed by the TRHBc theory microscopically because it includes self-consistently pairing correlations, continuum effects, and triaxial deformation degrees of freedom.
The deformation parameters $(\beta,\gamma)$ for $^{42}$Al are predicted to be $(0.35,42 ^\circ)$ by PC-PK1, $(0.36,42 ^\circ)$ by NL3*, and $(0.35,44 ^\circ)$ by NLSH.

Taking PC-PK1 as an example, the deformation parameters for $^{42}$Al are verified by the potential energy surface constructed from deformation constrained TRHBc calculations, as shown in Fig.~\ref{fig3}.
Without triaxiality, a prolate minimum at $\beta = 0.30$ and an oblate one at $\beta = -0.35$ are obtained.
With triaxiality, both the prolate and oblate minima are turned out to be saddle points.
After including the pairing correlations and the continuum effects, the existence of triaxiality in $^{42}$Al, the neutron-richest odd-odd aluminium isotope, provides an excellent platform to explore the halo phenomenon in triaxial nuclei.

\begin{figure}[htbp]
  \centering
  \includegraphics[width=0.4\textwidth]{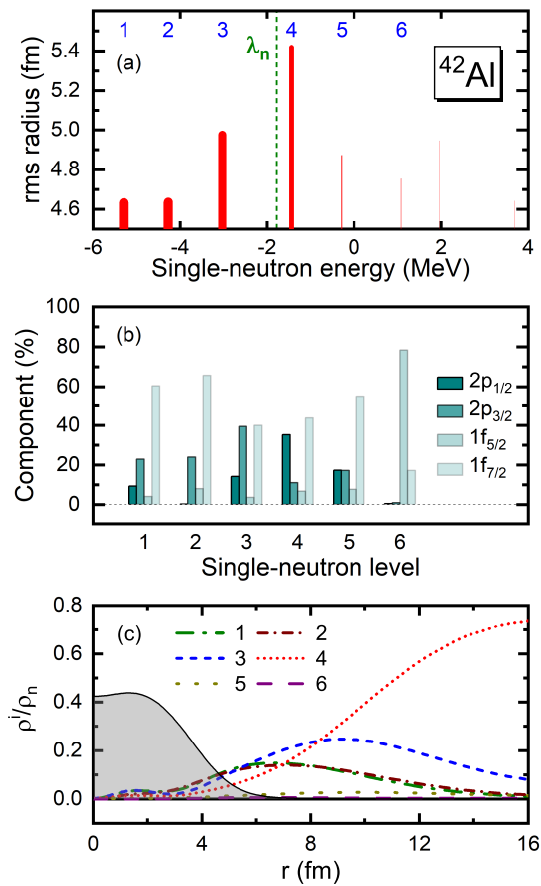}
  \caption{(a) The rms radius versus the energy $\epsilon$ for the single-neutron levels around the Fermi energy $\lambda_n$ in the canonical
                  basis for $^{42}$Al. The thickness of each level is proportional to its occupation probability. (b) The main spherical components for the single-neutron levels 1-6. (c) Contributions of the single-neutron levels to the total neutron density. The shaded region represents the total neutron density in arbitrary units. Here the angular dependence is averaged.}
\label{fig4}
\end{figure}

In order to examine the weakly bound levels and the continuum as well as their contributions in $^{42}$Al, in Fig.~\ref{fig4}, the single-neutron levels around the Fermi energy, their components, and their contributions to the total neutron density are shown.

In Fig.~\ref{fig4}(a), the rms radius is given versus the energy for the single-neutron levels around the Fermi energy in $^{42}$Al.
The thickness of each level is proportional to its occupation probability.
It is notable that the rms radius $5.4$~fm of level $4$, occupied by the last odd neutron, is significantly larger than those of its neighboring weakly-bound and continuum states.
This can be understood from the composition of level 4.

In Fig.~\ref{fig4}(b), the main components for the levels 1-6 are given.
The $35.3\%$ $2p_{1/2}$ and $11.1\%$ $2p_{3/2}$ components and the weak binding of level 4 account for its largest rms radius.
Although the contributions of $2p$ components are similar for levels $3$ and 4, the rms radius of level 3 is suppressed by its deeper binding.

In Fig.~\ref{fig4}(c), the contributions of the levels 1-6 to the total neutron density are shown as functions of radial coordinate $r$.
The contribution of the level $4$ becomes dominant after $r =9$~fm and even more than $70\%$ after $r = 14$~fm.
This can be understood from the low centrifugal barrier for $p$-wave components, which allows the considerable tunneling of the neutron into the classically forbidden region and the formation of neutron halo.

\begin{figure*}[htbp]
  \centering
  \includegraphics[width=0.8\textwidth]{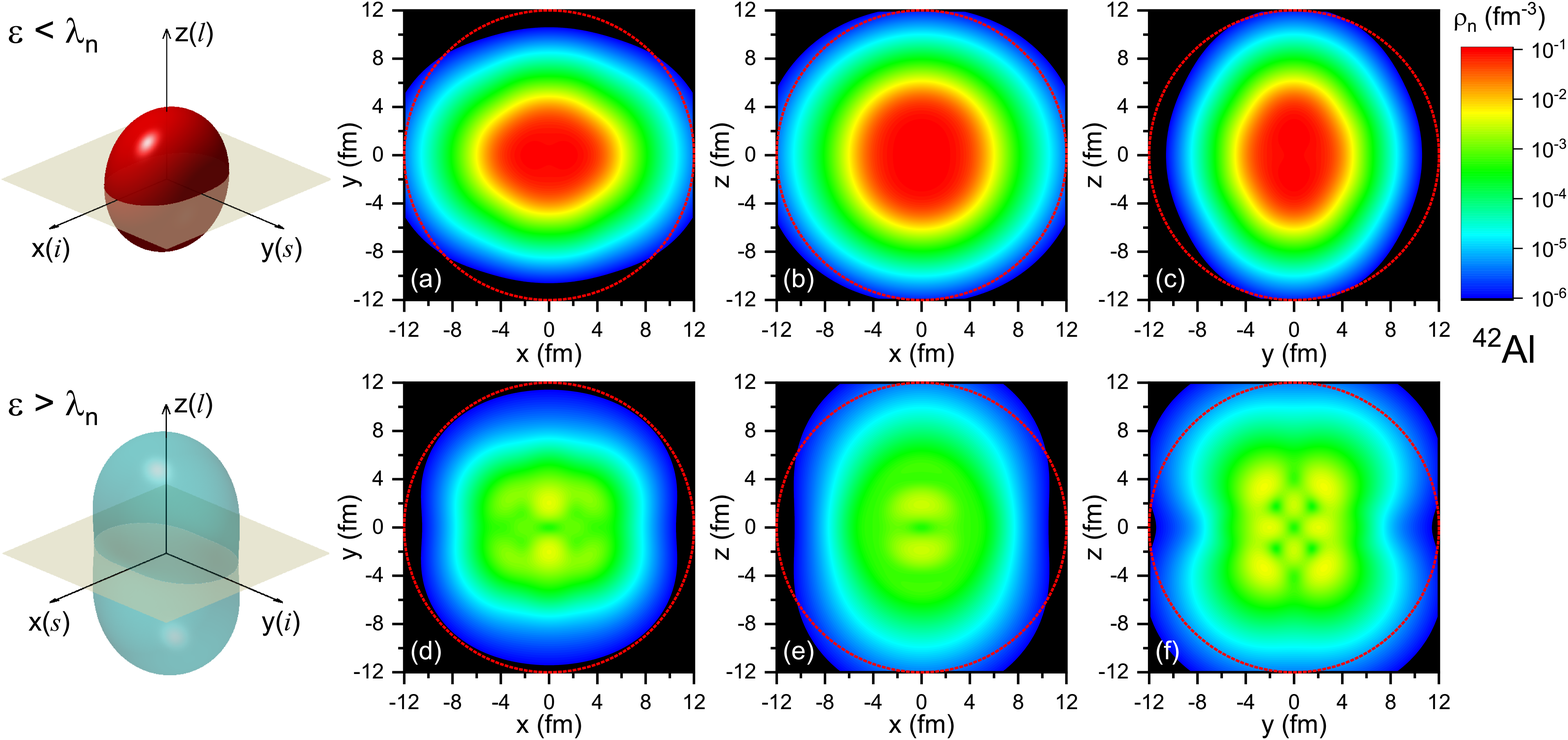}
  \caption{Neutron density distributions in $xy$, $xz$, and $yz$ planes contributed by the single-neutron levels with the energy $\epsilon$ below
                 and above the Fermi energy $\lambda_n$, i.e., $\epsilon<\lambda_n$ (a, b, c) and $\epsilon>\lambda_n$ (d, e, f). In each plot, a circle in dotted line is drawn to guide the eye. With the rms radius and deformation parameters $\beta$ and $\gamma$ from the densities with $\epsilon<\lambda_n$ and $\epsilon>\lambda_n$, the corresponding schematic shapes for upper and lower panels are given in the left, in which $s$, $i$, and $l$ respectively represent the short, intermediate, and long axes.}
\label{fig5}
\end{figure*}

The extended density distribution of level 4 and the energy gap of nearly 2~MeV between level 4 and level 3 provide a natural decomposition of the halo and the core.
This is equivalent with the use of the Fermi energy as a division.
Therefore, the neutron densities contributed by level $4$ and above ($\epsilon>\lambda_n$) and by level $3$ and below ($\epsilon<\lambda_n$) are shown in Fig.~\ref{fig5}.
Comparing the core and halo densities, the halo density does extend much farther than the core, particularly in the $yz$ plane, supporting a triaxially deformed one-neutron halo.
Quantitatively, the rms radii are $5.26$~fm for the halo and $3.85$~fm for the core.

The deformation parameters ($\beta$, $\gamma$) are ($0.79$, $-23^\circ$) for the halo and ($0.38$, $50^\circ$) for the core.
The negative $\gamma$ means that an exchange of the intermediate and short axes occurs for the halo and the core.
With the corresponding rms radius, $\beta$, and $\gamma$, schematic pictures are given in Fig.~\ref{fig5}, where the short, intermediate, and long axes can be clearly distinguished.
This shape decoupling between the core and the halo in $^{42}$Al includes the change of the deformation and the exchange of the intermediate and short axes.
It is even more exciting than the shape decoupling between the prolate core and the oblate halo predicted in $^{42,44}$Mg~\cite{Zhou2010PRC(R),Li2012PRC}.


In summary, a  microscopic self-consistent triaxial relativistic Hartree-Bogoliubov theory in continuum, which simultaneously takes into account the triaxiality and pairing correlations as well as continuum effects, is established and applied to explore the novel halo phenomenon in aluminum isotopes.
The experimental proton drip line, one-neutron separation energies, and charge radii are reproduced well by the TRHBc theory with PC-PK1, NL3*, NLSH, and PK1 density functionals, without any free parameters.
The observed $^{42}$Al is predicted to be the last bound odd-odd nucleus except for PK1.
The triaxial deformation in its ground state is verified by the constrained TRHBc calculations. 
The PC-PK1 predicted one-neutron separation energy is $0.68$ MeV, in excellent agreement with the AME2020 value of  $0.67(64)$ MeV, and the neutron rms radius is 3.94 fm, remarkably larger than the empirical value.
A new \emph{halo scale} $S_{\mathrm{halo}}$ is proposed to characterize the halo phenomenon in medium and heavy nuclei, and $^{42}$Al is turned out to be a one-neutron halo nucleus.
From the single-neutron levels around the Fermi energy, the valance neutron contributes dominantly to the neutron density at large $r$, due to its occupation of a weakly bound level with considerable $2p$ components.
The Fermi energy is found to be a natural division of the halo and the core for $^{42}$Al.
From the decomposed neutron density, novel shape decoupling between the core and the halo is found, i.e., the significant deformation change from $\beta=0.38$ to $0.79$ and the exchange of the intermediate and short axes with $\gamma$ from $50^\circ$ to $-23^\circ$.

Future experiments to explore the halo phenomenon and the novel shape decoupling in $^{42}$Al are highly demanded.
The nucleus $^{42}$Al was discovered in 2007, with a production rate of 1 in $10^{15}$ reactions~\cite{Baumann2007Nature}.
The measurement of the nuclear mass and radius can verify the weak binding and halo characters in $^{42}$Al.
Further experiments to explore the extended density distribution, $p$ components of the valance neutron, triaxial deformation, and shape decoupling are helpful to provide evidence for the halo phenomenon in triaxial nuclei.

\begin{acknowledgments}
Helpful discussions with L. S. Geng, P. Ring, B. H. Sun, D. Vretenar, P. W. Zhao, S.-G. Zhou, and members of the DRHBc Mass Table Collaboration are highly appreciated.
This work was partly supported by the National Natural Science Foundation of China (Grants No.~11935003, No.~11875075, No.~11975031, No.~12141501, and No.~12070131001), the National Key R\&D Program of China (Contract No.~2018YFA0404400), and High-performance Computing Platform of Peking University.
\end{acknowledgments}


\begin{thebibliography}{43}%
\makeatletter
\providecommand \@ifxundefined [1]{%
 \@ifx{#1\undefined}
}%
\providecommand \@ifnum [1]{%
 \ifnum #1\expandafter \@firstoftwo
 \else \expandafter \@secondoftwo
 \fi
}%
\providecommand \@ifx [1]{%
 \ifx #1\expandafter \@firstoftwo
 \else \expandafter \@secondoftwo
 \fi
}%
\providecommand \natexlab [1]{#1}%
\providecommand \enquote  [1]{``#1''}%
\providecommand \bibnamefont  [1]{#1}%
\providecommand \bibfnamefont [1]{#1}%
\providecommand \citenamefont [1]{#1}%
\providecommand \href@noop [0]{\@secondoftwo}%
\providecommand \href [0]{\begingroup \@sanitize@url \@href}%
\providecommand \@href[1]{\@@startlink{#1}\@@href}%
\providecommand \@@href[1]{\endgroup#1\@@endlink}%
\providecommand \@sanitize@url [0]{\catcode `\\12\catcode `\$12\catcode
  `\&12\catcode `\#12\catcode `\^12\catcode `\_12\catcode `\%12\relax}%
\providecommand \@@startlink[1]{}%
\providecommand \@@endlink[0]{}%
\providecommand \url  [0]{\begingroup\@sanitize@url \@url }%
\providecommand \@url [1]{\endgroup\@href {#1}{\urlprefix }}%
\providecommand \urlprefix  [0]{URL }%
\providecommand \Eprint [0]{\href }%
\providecommand \doibase [0]{http://dx.doi.org/}%
\providecommand \selectlanguage [0]{\@gobble}%
\providecommand \bibinfo  [0]{\@secondoftwo}%
\providecommand \bibfield  [0]{\@secondoftwo}%
\providecommand \translation [1]{[#1]}%
\providecommand \BibitemOpen [0]{}%
\providecommand \bibitemStop [0]{}%
\providecommand \bibitemNoStop [0]{.\EOS\space}%
\providecommand \EOS [0]{\spacefactor3000\relax}%
\providecommand \BibitemShut  [1]{\csname bibitem#1\endcsname}%
\let\auto@bib@innerbib\@empty
\bibitem [{\citenamefont {Jensen}\ \emph {et~al.}(2004)\citenamefont {Jensen},
  \citenamefont {Riisager}, \citenamefont {Fedorov},\ and\ \citenamefont
  {Garrido}}]{Jensen2004RMP}%
  \BibitemOpen
  \bibfield  {author} {\bibinfo {author} {\bibfnamefont {A.~S.}\ \bibnamefont
  {Jensen}}, \bibinfo {author} {\bibfnamefont {K.}~\bibnamefont {Riisager}},
  \bibinfo {author} {\bibfnamefont {D.~V.}\ \bibnamefont {Fedorov}}, \ and\
  \bibinfo {author} {\bibfnamefont {E.}~\bibnamefont {Garrido}},\ }\href
  {\doibase 10.1103/RevModPhys.76.215} {\bibfield  {journal} {\bibinfo
  {journal} {Rev. Mod. Phys.}\ }\textbf {\bibinfo {volume} {76}},\ \bibinfo
  {pages} {215} (\bibinfo {year} {2004})}\BibitemShut {NoStop}%
\bibitem [{\citenamefont {Tanihata}\ \emph {et~al.}(1985)\citenamefont
  {Tanihata}, \citenamefont {Hamagaki}, \citenamefont {Hashimoto},
  \citenamefont {Shida}, \citenamefont {Yoshikawa}, \citenamefont {Sugimoto},
  \citenamefont {Yamakawa}, \citenamefont {Kobayashi},\ and\ \citenamefont
  {Takahashi}}]{Tanihata1985PRL}%
  \BibitemOpen
  \bibfield  {author} {\bibinfo {author} {\bibfnamefont {I.}~\bibnamefont
  {Tanihata}}, \bibinfo {author} {\bibfnamefont {H.}~\bibnamefont {Hamagaki}},
  \bibinfo {author} {\bibfnamefont {O.}~\bibnamefont {Hashimoto}}, \bibinfo
  {author} {\bibfnamefont {Y.}~\bibnamefont {Shida}}, \bibinfo {author}
  {\bibfnamefont {N.}~\bibnamefont {Yoshikawa}}, \bibinfo {author}
  {\bibfnamefont {K.}~\bibnamefont {Sugimoto}}, \bibinfo {author}
  {\bibfnamefont {O.}~\bibnamefont {Yamakawa}}, \bibinfo {author}
  {\bibfnamefont {T.}~\bibnamefont {Kobayashi}}, \ and\ \bibinfo {author}
  {\bibfnamefont {N.}~\bibnamefont {Takahashi}},\ }\href {\doibase
  10.1103/PhysRevLett.55.2676} {\bibfield  {journal} {\bibinfo  {journal}
  {Phys. Rev. Lett.}\ }\textbf {\bibinfo {volume} {55}},\ \bibinfo {pages}
  {2676} (\bibinfo {year} {1985})}\BibitemShut {NoStop}%
\bibitem [{\citenamefont {Meng}\ and\ \citenamefont
  {Ring}(1996)}]{Meng1996PRL}%
  \BibitemOpen
  \bibfield  {author} {\bibinfo {author} {\bibfnamefont {J.}~\bibnamefont
  {Meng}}\ and\ \bibinfo {author} {\bibfnamefont {P.}~\bibnamefont {Ring}},\
  }\href {\doibase 10.1103/PhysRevLett.77.3963} {\bibfield  {journal} {\bibinfo
   {journal} {Phys. Rev. Lett.}\ }\textbf {\bibinfo {volume} {77}},\ \bibinfo
  {pages} {3963} (\bibinfo {year} {1996})}\BibitemShut {NoStop}%
\bibitem [{\citenamefont {Meng}\ and\ \citenamefont
  {Ring}(1998)}]{Meng1998PRL}%
  \BibitemOpen
  \bibfield  {author} {\bibinfo {author} {\bibfnamefont {J.}~\bibnamefont
  {Meng}}\ and\ \bibinfo {author} {\bibfnamefont {P.}~\bibnamefont {Ring}},\
  }\href {\doibase 10.1103/PhysRevLett.80.460} {\bibfield  {journal} {\bibinfo
  {journal} {Phys. Rev. Lett.}\ }\textbf {\bibinfo {volume} {80}},\ \bibinfo
  {pages} {460} (\bibinfo {year} {1998})}\BibitemShut {NoStop}%
\bibitem [{\citenamefont {Tanihata}\ \emph {et~al.}(1995)\citenamefont
  {Tanihata}, \citenamefont {Hirata},\ and\ \citenamefont
  {Toki}}]{Tanihata1995NPA}%
  \BibitemOpen
  \bibfield  {author} {\bibinfo {author} {\bibfnamefont {I.}~\bibnamefont
  {Tanihata}}, \bibinfo {author} {\bibfnamefont {D.}~\bibnamefont {Hirata}}, \
  and\ \bibinfo {author} {\bibfnamefont {H.}~\bibnamefont {Toki}},\ }\href
  {\doibase 10.1016/0375-9474(94)00756-D} {\bibfield  {journal} {\bibinfo
  {journal} {Nucl. Phys. A}\ }\textbf {\bibinfo {volume} {583}},\ \bibinfo
  {pages} {769} (\bibinfo {year} {1995})}\BibitemShut {NoStop}%
\bibitem [{\citenamefont {Hamamoto}(2004)}]{Hamamoto2004PRC(R)}%
  \BibitemOpen
  \bibfield  {author} {\bibinfo {author} {\bibfnamefont {I.}~\bibnamefont
  {Hamamoto}},\ }\href {\doibase 10.1103/PhysRevC.69.041306} {\bibfield
  {journal} {\bibinfo  {journal} {Phys. Rev. C}\ }\textbf {\bibinfo {volume}
  {69}},\ \bibinfo {pages} {041306(R)} (\bibinfo {year} {2004})}\BibitemShut
  {NoStop}%
\bibitem [{\citenamefont {Nunes}(2005)}]{Nunes2005NPA}%
  \BibitemOpen
  \bibfield  {author} {\bibinfo {author} {\bibfnamefont {F.}~\bibnamefont
  {Nunes}},\ }\href {\doibase 10.1016/j.nuclphysa.2005.04.005} {\bibfield
  {journal} {\bibinfo  {journal} {Nucl. Phys. A}\ }\textbf {\bibinfo {volume}
  {757}},\ \bibinfo {pages} {349 } (\bibinfo {year} {2005})}\BibitemShut
  {NoStop}%
\bibitem [{\citenamefont {Zhou}\ \emph {et~al.}(2010)\citenamefont {Zhou},
  \citenamefont {Meng}, \citenamefont {Ring},\ and\ \citenamefont
  {Zhao}}]{Zhou2010PRC(R)}%
  \BibitemOpen
  \bibfield  {author} {\bibinfo {author} {\bibfnamefont {S.-G.}\ \bibnamefont
  {Zhou}}, \bibinfo {author} {\bibfnamefont {J.}~\bibnamefont {Meng}}, \bibinfo
  {author} {\bibfnamefont {P.}~\bibnamefont {Ring}}, \ and\ \bibinfo {author}
  {\bibfnamefont {E.-G.}\ \bibnamefont {Zhao}},\ }\href {\doibase
  10.1103/PhysRevC.82.011301} {\bibfield  {journal} {\bibinfo  {journal} {Phys.
  Rev. C}\ }\textbf {\bibinfo {volume} {82}},\ \bibinfo {pages} {011301(R)}
  (\bibinfo {year} {2010})}\BibitemShut {NoStop}%
\bibitem [{\citenamefont {Li}\ \emph {et~al.}(2012{\natexlab{a}})\citenamefont
  {Li}, \citenamefont {Meng}, \citenamefont {Ring}, \citenamefont {Zhao},\ and\
  \citenamefont {Zhou}}]{Li2012PRC}%
  \BibitemOpen
  \bibfield  {author} {\bibinfo {author} {\bibfnamefont {L.}~\bibnamefont
  {Li}}, \bibinfo {author} {\bibfnamefont {J.}~\bibnamefont {Meng}}, \bibinfo
  {author} {\bibfnamefont {P.}~\bibnamefont {Ring}}, \bibinfo {author}
  {\bibfnamefont {E.-G.}\ \bibnamefont {Zhao}}, \ and\ \bibinfo {author}
  {\bibfnamefont {S.-G.}\ \bibnamefont {Zhou}},\ }\href {\doibase
  10.1103/PhysRevC.85.024312} {\bibfield  {journal} {\bibinfo  {journal} {Phys.
  Rev. C}\ }\textbf {\bibinfo {volume} {85}},\ \bibinfo {pages} {024312}
  (\bibinfo {year} {2012}{\natexlab{a}})}\BibitemShut {NoStop}%
\bibitem [{\citenamefont {Nakamura}\ \emph {et~al.}(2014)\citenamefont
  {Nakamura}, \citenamefont {Kobayashi}, \citenamefont {Kondo}, \citenamefont
  {Satou}, \citenamefont {Tostevin}, \citenamefont {Utsuno}, \citenamefont
  {Aoi}, \citenamefont {Baba}, \citenamefont {Fukuda}, \citenamefont {Gibelin},
  \citenamefont {Inabe}, \citenamefont {Ishihara}, \citenamefont {Kameda},
  \citenamefont {Kubo}, \citenamefont {Motobayashi}, \citenamefont {Ohnishi},
  \citenamefont {Orr}, \citenamefont {Otsu}, \citenamefont {Otsuka},
  \citenamefont {Sakurai}, \citenamefont {Sumikama}, \citenamefont {Takeda},
  \citenamefont {Takeshita}, \citenamefont {Takechi}, \citenamefont {Takeuchi},
  \citenamefont {Togano},\ and\ \citenamefont {Yoneda}}]{Nakamura2014PRL}%
  \BibitemOpen
  \bibfield  {author} {\bibinfo {author} {\bibfnamefont {T.}~\bibnamefont
  {Nakamura}}, \bibinfo {author} {\bibfnamefont {N.}~\bibnamefont {Kobayashi}},
  \bibinfo {author} {\bibfnamefont {Y.}~\bibnamefont {Kondo}}, \bibinfo
  {author} {\bibfnamefont {Y.}~\bibnamefont {Satou}}, \bibinfo {author}
  {\bibfnamefont {J.~A.}\ \bibnamefont {Tostevin}}, \bibinfo {author}
  {\bibfnamefont {Y.}~\bibnamefont {Utsuno}}, \bibinfo {author} {\bibfnamefont
  {N.}~\bibnamefont {Aoi}}, \bibinfo {author} {\bibfnamefont {H.}~\bibnamefont
  {Baba}}, \bibinfo {author} {\bibfnamefont {N.}~\bibnamefont {Fukuda}},
  \bibinfo {author} {\bibfnamefont {J.}~\bibnamefont {Gibelin}}, \bibinfo
  {author} {\bibfnamefont {N.}~\bibnamefont {Inabe}}, \bibinfo {author}
  {\bibfnamefont {M.}~\bibnamefont {Ishihara}}, \bibinfo {author}
  {\bibfnamefont {D.}~\bibnamefont {Kameda}}, \bibinfo {author} {\bibfnamefont
  {T.}~\bibnamefont {Kubo}}, \bibinfo {author} {\bibfnamefont {T.}~\bibnamefont
  {Motobayashi}}, \bibinfo {author} {\bibfnamefont {T.}~\bibnamefont
  {Ohnishi}}, \bibinfo {author} {\bibfnamefont {N.~A.}\ \bibnamefont {Orr}},
  \bibinfo {author} {\bibfnamefont {H.}~\bibnamefont {Otsu}}, \bibinfo {author}
  {\bibfnamefont {T.}~\bibnamefont {Otsuka}}, \bibinfo {author} {\bibfnamefont
  {H.}~\bibnamefont {Sakurai}}, \bibinfo {author} {\bibfnamefont
  {T.}~\bibnamefont {Sumikama}}, \bibinfo {author} {\bibfnamefont
  {H.}~\bibnamefont {Takeda}}, \bibinfo {author} {\bibfnamefont
  {E.}~\bibnamefont {Takeshita}}, \bibinfo {author} {\bibfnamefont
  {M.}~\bibnamefont {Takechi}}, \bibinfo {author} {\bibfnamefont
  {S.}~\bibnamefont {Takeuchi}}, \bibinfo {author} {\bibfnamefont
  {Y.}~\bibnamefont {Togano}}, \ and\ \bibinfo {author} {\bibfnamefont
  {K.}~\bibnamefont {Yoneda}},\ }\href {\doibase
  10.1103/PhysRevLett.112.142501} {\bibfield  {journal} {\bibinfo  {journal}
  {Phys. Rev. Lett.}\ }\textbf {\bibinfo {volume} {112}},\ \bibinfo {pages}
  {142501} (\bibinfo {year} {2014})}\BibitemShut {NoStop}%
\bibitem [{\citenamefont {Kobayashi}\ \emph {et~al.}(2014)\citenamefont
  {Kobayashi}, \citenamefont {Nakamura}, \citenamefont {Kondo}, \citenamefont
  {Tostevin}, \citenamefont {Utsuno}, \citenamefont {Aoi}, \citenamefont
  {Baba}, \citenamefont {Barthelemy}, \citenamefont {Famiano}, \citenamefont
  {Fukuda}, \citenamefont {Inabe}, \citenamefont {Ishihara}, \citenamefont
  {Kanungo}, \citenamefont {Kim}, \citenamefont {Kubo}, \citenamefont {Lee},
  \citenamefont {Lee}, \citenamefont {Matsushita}, \citenamefont {Motobayashi},
  \citenamefont {Ohnishi}, \citenamefont {Orr}, \citenamefont {Otsu},
  \citenamefont {Otsuka}, \citenamefont {Sako}, \citenamefont {Sakurai},
  \citenamefont {Satou}, \citenamefont {Sumikama}, \citenamefont {Takeda},
  \citenamefont {Takeuchi}, \citenamefont {Tanaka}, \citenamefont {Togano},\
  and\ \citenamefont {Yoneda}}]{Kobayashi2014PRL}%
  \BibitemOpen
  \bibfield  {author} {\bibinfo {author} {\bibfnamefont {N.}~\bibnamefont
  {Kobayashi}}, \bibinfo {author} {\bibfnamefont {T.}~\bibnamefont {Nakamura}},
  \bibinfo {author} {\bibfnamefont {Y.}~\bibnamefont {Kondo}}, \bibinfo
  {author} {\bibfnamefont {J.~A.}\ \bibnamefont {Tostevin}}, \bibinfo {author}
  {\bibfnamefont {Y.}~\bibnamefont {Utsuno}}, \bibinfo {author} {\bibfnamefont
  {N.}~\bibnamefont {Aoi}}, \bibinfo {author} {\bibfnamefont {H.}~\bibnamefont
  {Baba}}, \bibinfo {author} {\bibfnamefont {R.}~\bibnamefont {Barthelemy}},
  \bibinfo {author} {\bibfnamefont {M.~A.}\ \bibnamefont {Famiano}}, \bibinfo
  {author} {\bibfnamefont {N.}~\bibnamefont {Fukuda}}, \bibinfo {author}
  {\bibfnamefont {N.}~\bibnamefont {Inabe}}, \bibinfo {author} {\bibfnamefont
  {M.}~\bibnamefont {Ishihara}}, \bibinfo {author} {\bibfnamefont
  {R.}~\bibnamefont {Kanungo}}, \bibinfo {author} {\bibfnamefont
  {S.}~\bibnamefont {Kim}}, \bibinfo {author} {\bibfnamefont {T.}~\bibnamefont
  {Kubo}}, \bibinfo {author} {\bibfnamefont {G.~S.}\ \bibnamefont {Lee}},
  \bibinfo {author} {\bibfnamefont {H.~S.}\ \bibnamefont {Lee}}, \bibinfo
  {author} {\bibfnamefont {M.}~\bibnamefont {Matsushita}}, \bibinfo {author}
  {\bibfnamefont {T.}~\bibnamefont {Motobayashi}}, \bibinfo {author}
  {\bibfnamefont {T.}~\bibnamefont {Ohnishi}}, \bibinfo {author} {\bibfnamefont
  {N.~A.}\ \bibnamefont {Orr}}, \bibinfo {author} {\bibfnamefont
  {H.}~\bibnamefont {Otsu}}, \bibinfo {author} {\bibfnamefont {T.}~\bibnamefont
  {Otsuka}}, \bibinfo {author} {\bibfnamefont {T.}~\bibnamefont {Sako}},
  \bibinfo {author} {\bibfnamefont {H.}~\bibnamefont {Sakurai}}, \bibinfo
  {author} {\bibfnamefont {Y.}~\bibnamefont {Satou}}, \bibinfo {author}
  {\bibfnamefont {T.}~\bibnamefont {Sumikama}}, \bibinfo {author}
  {\bibfnamefont {H.}~\bibnamefont {Takeda}}, \bibinfo {author} {\bibfnamefont
  {S.}~\bibnamefont {Takeuchi}}, \bibinfo {author} {\bibfnamefont
  {R.}~\bibnamefont {Tanaka}}, \bibinfo {author} {\bibfnamefont
  {Y.}~\bibnamefont {Togano}}, \ and\ \bibinfo {author} {\bibfnamefont
  {K.}~\bibnamefont {Yoneda}},\ }\href {\doibase
  10.1103/PhysRevLett.112.242501} {\bibfield  {journal} {\bibinfo  {journal}
  {Phys. Rev. Lett.}\ }\textbf {\bibinfo {volume} {112}},\ \bibinfo {pages}
  {242501} (\bibinfo {year} {2014})}\BibitemShut {NoStop}%
\bibitem [{\citenamefont {Yang}\ \emph {et~al.}(2021)\citenamefont {Yang},
  \citenamefont {Kubota}, \citenamefont {Corsi}, \citenamefont {Yoshida},
  \citenamefont {Sun}, \citenamefont {Li}, \citenamefont {Kimura},
  \citenamefont {Michel}, \citenamefont {Ogata}, \citenamefont {Yuan},
  \citenamefont {Yuan}, \citenamefont {Authelet}, \citenamefont {Baba},
  \citenamefont {Caesar}, \citenamefont {Calvet}, \citenamefont {Delbart},
  \citenamefont {Dozono}, \citenamefont {Feng}, \citenamefont {Flavigny},
  \citenamefont {Gheller}, \citenamefont {Gibelin}, \citenamefont {Giganon},
  \citenamefont {Gillibert}, \citenamefont {Hasegawa}, \citenamefont {Isobe},
  \citenamefont {Kanaya}, \citenamefont {Kawakami}, \citenamefont {Kim},
  \citenamefont {Kiyokawa}, \citenamefont {Kobayashi}, \citenamefont
  {Kobayashi}, \citenamefont {Kobayashi}, \citenamefont {Kondo}, \citenamefont
  {Korkulu}, \citenamefont {Koyama}, \citenamefont {Lapoux}, \citenamefont
  {Maeda}, \citenamefont {Marqu\'es}, \citenamefont {Motobayashi},
  \citenamefont {Miyazaki}, \citenamefont {Nakamura}, \citenamefont
  {Nakatsuka}, \citenamefont {Nishio}, \citenamefont {Obertelli}, \citenamefont
  {Ohkura}, \citenamefont {Orr}, \citenamefont {Ota}, \citenamefont {Otsu},
  \citenamefont {Ozaki}, \citenamefont {Panin}, \citenamefont {Paschalis},
  \citenamefont {Pollacco}, \citenamefont {Reichert}, \citenamefont {Rouss\'e},
  \citenamefont {Saito}, \citenamefont {Sakaguchi}, \citenamefont {Sako},
  \citenamefont {Santamaria}, \citenamefont {Sasano}, \citenamefont {Sato},
  \citenamefont {Shikata}, \citenamefont {Shimizu}, \citenamefont {Shindo},
  \citenamefont {Stuhl}, \citenamefont {Sumikama}, \citenamefont {Sun},
  \citenamefont {Tabata}, \citenamefont {Togano}, \citenamefont {Tsubota},
  \citenamefont {Xu}, \citenamefont {Yasuda}, \citenamefont {Yoneda},
  \citenamefont {Zenihiro}, \citenamefont {Zhou}, \citenamefont {Zuo},\ and\
  \citenamefont {Uesaka}}]{Yang2021PRL}%
  \BibitemOpen
  \bibfield  {author} {\bibinfo {author} {\bibfnamefont {Z.~H.}\ \bibnamefont
  {Yang}}, \bibinfo {author} {\bibfnamefont {Y.}~\bibnamefont {Kubota}},
  \bibinfo {author} {\bibfnamefont {A.}~\bibnamefont {Corsi}}, \bibinfo
  {author} {\bibfnamefont {K.}~\bibnamefont {Yoshida}}, \bibinfo {author}
  {\bibfnamefont {X.-X.}\ \bibnamefont {Sun}}, \bibinfo {author} {\bibfnamefont
  {J.~G.}\ \bibnamefont {Li}}, \bibinfo {author} {\bibfnamefont
  {M.}~\bibnamefont {Kimura}}, \bibinfo {author} {\bibfnamefont
  {N.}~\bibnamefont {Michel}}, \bibinfo {author} {\bibfnamefont
  {K.}~\bibnamefont {Ogata}}, \bibinfo {author} {\bibfnamefont {C.~X.}\
  \bibnamefont {Yuan}}, \bibinfo {author} {\bibfnamefont {Q.}~\bibnamefont
  {Yuan}}, \bibinfo {author} {\bibfnamefont {G.}~\bibnamefont {Authelet}},
  \bibinfo {author} {\bibfnamefont {H.}~\bibnamefont {Baba}}, \bibinfo {author}
  {\bibfnamefont {C.}~\bibnamefont {Caesar}}, \bibinfo {author} {\bibfnamefont
  {D.}~\bibnamefont {Calvet}}, \bibinfo {author} {\bibfnamefont
  {A.}~\bibnamefont {Delbart}}, \bibinfo {author} {\bibfnamefont
  {M.}~\bibnamefont {Dozono}}, \bibinfo {author} {\bibfnamefont
  {J.}~\bibnamefont {Feng}}, \bibinfo {author} {\bibfnamefont {F.}~\bibnamefont
  {Flavigny}}, \bibinfo {author} {\bibfnamefont {J.-M.}\ \bibnamefont
  {Gheller}}, \bibinfo {author} {\bibfnamefont {J.}~\bibnamefont {Gibelin}},
  \bibinfo {author} {\bibfnamefont {A.}~\bibnamefont {Giganon}}, \bibinfo
  {author} {\bibfnamefont {A.}~\bibnamefont {Gillibert}}, \bibinfo {author}
  {\bibfnamefont {K.}~\bibnamefont {Hasegawa}}, \bibinfo {author}
  {\bibfnamefont {T.}~\bibnamefont {Isobe}}, \bibinfo {author} {\bibfnamefont
  {Y.}~\bibnamefont {Kanaya}}, \bibinfo {author} {\bibfnamefont
  {S.}~\bibnamefont {Kawakami}}, \bibinfo {author} {\bibfnamefont
  {D.}~\bibnamefont {Kim}}, \bibinfo {author} {\bibfnamefont {Y.}~\bibnamefont
  {Kiyokawa}}, \bibinfo {author} {\bibfnamefont {M.}~\bibnamefont {Kobayashi}},
  \bibinfo {author} {\bibfnamefont {N.}~\bibnamefont {Kobayashi}}, \bibinfo
  {author} {\bibfnamefont {T.}~\bibnamefont {Kobayashi}}, \bibinfo {author}
  {\bibfnamefont {Y.}~\bibnamefont {Kondo}}, \bibinfo {author} {\bibfnamefont
  {Z.}~\bibnamefont {Korkulu}}, \bibinfo {author} {\bibfnamefont
  {S.}~\bibnamefont {Koyama}}, \bibinfo {author} {\bibfnamefont
  {V.}~\bibnamefont {Lapoux}}, \bibinfo {author} {\bibfnamefont
  {Y.}~\bibnamefont {Maeda}}, \bibinfo {author} {\bibfnamefont {F.~M.}\
  \bibnamefont {Marqu\'es}}, \bibinfo {author} {\bibfnamefont {T.}~\bibnamefont
  {Motobayashi}}, \bibinfo {author} {\bibfnamefont {T.}~\bibnamefont
  {Miyazaki}}, \bibinfo {author} {\bibfnamefont {T.}~\bibnamefont {Nakamura}},
  \bibinfo {author} {\bibfnamefont {N.}~\bibnamefont {Nakatsuka}}, \bibinfo
  {author} {\bibfnamefont {Y.}~\bibnamefont {Nishio}}, \bibinfo {author}
  {\bibfnamefont {A.}~\bibnamefont {Obertelli}}, \bibinfo {author}
  {\bibfnamefont {A.}~\bibnamefont {Ohkura}}, \bibinfo {author} {\bibfnamefont
  {N.~A.}\ \bibnamefont {Orr}}, \bibinfo {author} {\bibfnamefont
  {S.}~\bibnamefont {Ota}}, \bibinfo {author} {\bibfnamefont {H.}~\bibnamefont
  {Otsu}}, \bibinfo {author} {\bibfnamefont {T.}~\bibnamefont {Ozaki}},
  \bibinfo {author} {\bibfnamefont {V.}~\bibnamefont {Panin}}, \bibinfo
  {author} {\bibfnamefont {S.}~\bibnamefont {Paschalis}}, \bibinfo {author}
  {\bibfnamefont {E.~C.}\ \bibnamefont {Pollacco}}, \bibinfo {author}
  {\bibfnamefont {S.}~\bibnamefont {Reichert}}, \bibinfo {author}
  {\bibfnamefont {J.-Y.}\ \bibnamefont {Rouss\'e}}, \bibinfo {author}
  {\bibfnamefont {A.~T.}\ \bibnamefont {Saito}}, \bibinfo {author}
  {\bibfnamefont {S.}~\bibnamefont {Sakaguchi}}, \bibinfo {author}
  {\bibfnamefont {M.}~\bibnamefont {Sako}}, \bibinfo {author} {\bibfnamefont
  {C.}~\bibnamefont {Santamaria}}, \bibinfo {author} {\bibfnamefont
  {M.}~\bibnamefont {Sasano}}, \bibinfo {author} {\bibfnamefont
  {H.}~\bibnamefont {Sato}}, \bibinfo {author} {\bibfnamefont {M.}~\bibnamefont
  {Shikata}}, \bibinfo {author} {\bibfnamefont {Y.}~\bibnamefont {Shimizu}},
  \bibinfo {author} {\bibfnamefont {Y.}~\bibnamefont {Shindo}}, \bibinfo
  {author} {\bibfnamefont {L.}~\bibnamefont {Stuhl}}, \bibinfo {author}
  {\bibfnamefont {T.}~\bibnamefont {Sumikama}}, \bibinfo {author}
  {\bibfnamefont {Y.~L.}\ \bibnamefont {Sun}}, \bibinfo {author} {\bibfnamefont
  {M.}~\bibnamefont {Tabata}}, \bibinfo {author} {\bibfnamefont
  {Y.}~\bibnamefont {Togano}}, \bibinfo {author} {\bibfnamefont
  {J.}~\bibnamefont {Tsubota}}, \bibinfo {author} {\bibfnamefont {F.~R.}\
  \bibnamefont {Xu}}, \bibinfo {author} {\bibfnamefont {J.}~\bibnamefont
  {Yasuda}}, \bibinfo {author} {\bibfnamefont {K.}~\bibnamefont {Yoneda}},
  \bibinfo {author} {\bibfnamefont {J.}~\bibnamefont {Zenihiro}}, \bibinfo
  {author} {\bibfnamefont {S.-G.}\ \bibnamefont {Zhou}}, \bibinfo {author}
  {\bibfnamefont {W.}~\bibnamefont {Zuo}}, \ and\ \bibinfo {author}
  {\bibfnamefont {T.}~\bibnamefont {Uesaka}},\ }\href {\doibase
  10.1103/PhysRevLett.126.082501} {\bibfield  {journal} {\bibinfo  {journal}
  {Phys. Rev. Lett.}\ }\textbf {\bibinfo {volume} {126}},\ \bibinfo {pages}
  {082501} (\bibinfo {year} {2021})}\BibitemShut {NoStop}%
\bibitem [{\citenamefont {Sun}(2021)}]{Sun2021PRC(1)}%
  \BibitemOpen
  \bibfield  {author} {\bibinfo {author} {\bibfnamefont {X.-X.}\ \bibnamefont
  {Sun}},\ }\href {\doibase 10.1103/PhysRevC.103.054315} {\bibfield  {journal}
  {\bibinfo  {journal} {Phys. Rev. C}\ }\textbf {\bibinfo {volume} {103}},\
  \bibinfo {pages} {054315} (\bibinfo {year} {2021})}\BibitemShut {NoStop}%
\bibitem [{\citenamefont {Sun}\ \emph {et~al.}(2018)\citenamefont {Sun},
  \citenamefont {Zhao},\ and\ \citenamefont {Zhou}}]{Sun2018PLB}%
  \BibitemOpen
  \bibfield  {author} {\bibinfo {author} {\bibfnamefont {X.-X.}\ \bibnamefont
  {Sun}}, \bibinfo {author} {\bibfnamefont {J.}~\bibnamefont {Zhao}}, \ and\
  \bibinfo {author} {\bibfnamefont {S.-G.}\ \bibnamefont {Zhou}},\ }\href
  {\doibase 10.1016/j.physletb.2018.08.071} {\bibfield  {journal} {\bibinfo
  {journal} {Phys. Lett. B}\ }\textbf {\bibinfo {volume} {785}},\ \bibinfo
  {pages} {530} (\bibinfo {year} {2018})}\BibitemShut {NoStop}%
\bibitem [{\citenamefont {Sun}\ \emph {et~al.}(2020)\citenamefont {Sun},
  \citenamefont {Zhao},\ and\ \citenamefont {Zhou}}]{Sun2020NPA}%
  \BibitemOpen
  \bibfield  {author} {\bibinfo {author} {\bibfnamefont {X.-X.}\ \bibnamefont
  {Sun}}, \bibinfo {author} {\bibfnamefont {J.}~\bibnamefont {Zhao}}, \ and\
  \bibinfo {author} {\bibfnamefont {S.-G.}\ \bibnamefont {Zhou}},\ }\href
  {\doibase 10.1016/j.nuclphysa.2020.122011} {\bibfield  {journal} {\bibinfo
  {journal} {Nucl. Phys. A}\ }\textbf {\bibinfo {volume} {1003}},\ \bibinfo
  {pages} {122011} (\bibinfo {year} {2020})}\BibitemShut {NoStop}%
\bibitem [{\citenamefont {Zhong}\ \emph {et~al.}(2022)\citenamefont {Zhong},
  \citenamefont {Zhang}, \citenamefont {Sun},\ and\ \citenamefont
  {Smith}}]{Zhong2022SciChina}%
  \BibitemOpen
  \bibfield  {author} {\bibinfo {author} {\bibfnamefont {S.~Y.}\ \bibnamefont
  {Zhong}}, \bibinfo {author} {\bibfnamefont {S.~S.}\ \bibnamefont {Zhang}},
  \bibinfo {author} {\bibfnamefont {X.~X.}\ \bibnamefont {Sun}}, \ and\
  \bibinfo {author} {\bibfnamefont {M.~S.}\ \bibnamefont {Smith}},\ }\href
  {https://www.sciengine.com/SCPMA/article?doi=10.1007/s11433-022-1894-6&scroll=}
  {\bibfield  {journal} {\bibinfo  {journal} {Sci. China Phys. Mech. Astron.}\
  }\textbf {\bibinfo {volume} {65}},\ \bibinfo {pages} {262011} (\bibinfo
  {year} {2022})}\BibitemShut {NoStop}%
\bibitem [{\citenamefont {Zhang}\ \emph {et~al.}(2019)\citenamefont {Zhang},
  \citenamefont {Wang},\ and\ \citenamefont {Zhang}}]{Zhang2019PRC}%
  \BibitemOpen
  \bibfield  {author} {\bibinfo {author} {\bibfnamefont {K.~Y.}\ \bibnamefont
  {Zhang}}, \bibinfo {author} {\bibfnamefont {D.~Y.}\ \bibnamefont {Wang}}, \
  and\ \bibinfo {author} {\bibfnamefont {S.~Q.}\ \bibnamefont {Zhang}},\ }\href
  {\doibase 10.1103/PhysRevC.100.034312} {\bibfield  {journal} {\bibinfo
  {journal} {Phys. Rev. C}\ }\textbf {\bibinfo {volume} {100}},\ \bibinfo
  {pages} {034312} (\bibinfo {year} {2019})}\BibitemShut {NoStop}%
\bibitem [{\citenamefont {Sun}\ and\ \citenamefont
  {Zhou}(2021)}]{Sun2021SciBull}%
  \BibitemOpen
  \bibfield  {author} {\bibinfo {author} {\bibfnamefont {X.-X.}\ \bibnamefont
  {Sun}}\ and\ \bibinfo {author} {\bibfnamefont {S.-G.}\ \bibnamefont {Zhou}},\
  }\href {\doibase 10.1016/j.scib.2021.07.005} {\bibfield  {journal} {\bibinfo
  {journal} {Sci. Bull.}\ }\textbf {\bibinfo {volume} {66}},\ \bibinfo {pages}
  {2072} (\bibinfo {year} {2021})}\BibitemShut {NoStop}%
\bibitem [{\citenamefont {Lu}\ \emph {et~al.}(2012)\citenamefont {Lu},
  \citenamefont {Zhao},\ and\ \citenamefont {Zhou}}]{Lyu2012PRC(R)}%
  \BibitemOpen
  \bibfield  {author} {\bibinfo {author} {\bibfnamefont {B.-N.}\ \bibnamefont
  {Lu}}, \bibinfo {author} {\bibfnamefont {E.-G.}\ \bibnamefont {Zhao}}, \ and\
  \bibinfo {author} {\bibfnamefont {S.-G.}\ \bibnamefont {Zhou}},\ }\href
  {\doibase 10.1103/PhysRevC.85.011301} {\bibfield  {journal} {\bibinfo
  {journal} {Phys. Rev. C}\ }\textbf {\bibinfo {volume} {85}},\ \bibinfo
  {pages} {011301(R)} (\bibinfo {year} {2012})}\BibitemShut {NoStop}%
\bibitem [{\citenamefont {Frauendorf}\ and\ \citenamefont
  {Meng}(1997)}]{Frauendorf1997NPA}%
  \BibitemOpen
  \bibfield  {author} {\bibinfo {author} {\bibfnamefont {S.}~\bibnamefont
  {Frauendorf}}\ and\ \bibinfo {author} {\bibfnamefont {J.}~\bibnamefont
  {Meng}},\ }\href {\doibase 10.1016/S0375-9474(97)00004-3} {\bibfield
  {journal} {\bibinfo  {journal} {Nucl. Phys. A}\ }\textbf {\bibinfo {volume}
  {617}},\ \bibinfo {pages} {131} (\bibinfo {year} {1997})}\BibitemShut
  {NoStop}%
\bibitem [{\citenamefont {Bohr}\ and\ \citenamefont
  {Mottelson}(1975)}]{Bohr1975Book}%
  \BibitemOpen
  \bibfield  {author} {\bibinfo {author} {\bibfnamefont {A.}~\bibnamefont
  {Bohr}}\ and\ \bibinfo {author} {\bibfnamefont {B.~R.}\ \bibnamefont
  {Mottelson}},\ }\href@noop {} {\emph {\bibinfo {title} {Nuclear
  Structure}}},\ Vol.~\bibinfo {volume} {II}\ (\bibinfo  {publisher} {Benjamin,
  New York},\ \bibinfo {year} {1975})\BibitemShut {NoStop}%
\bibitem [{\citenamefont {Uzawa}\ \emph {et~al.}(2021)\citenamefont {Uzawa},
  \citenamefont {Hagino},\ and\ \citenamefont {Yoshida}}]{Uzawa2021PRC(L)}%
  \BibitemOpen
  \bibfield  {author} {\bibinfo {author} {\bibfnamefont {K.}~\bibnamefont
  {Uzawa}}, \bibinfo {author} {\bibfnamefont {K.}~\bibnamefont {Hagino}}, \
  and\ \bibinfo {author} {\bibfnamefont {K.}~\bibnamefont {Yoshida}},\ }\href
  {\doibase 10.1103/PhysRevC.104.L011303} {\bibfield  {journal} {\bibinfo
  {journal} {Phys. Rev. C}\ }\textbf {\bibinfo {volume} {104}},\ \bibinfo
  {pages} {L011303} (\bibinfo {year} {2021})}\BibitemShut {NoStop}%
\bibitem [{\citenamefont {Kucharek}\ and\ \citenamefont
  {Ring}(1991)}]{Kucharek1991ZPA}%
  \BibitemOpen
  \bibfield  {author} {\bibinfo {author} {\bibfnamefont {H.}~\bibnamefont
  {Kucharek}}\ and\ \bibinfo {author} {\bibfnamefont {P.}~\bibnamefont
  {Ring}},\ }\href
  {https://link.springer.com/content/pdf/10.1007/BF01282930.pdf} {\bibfield
  {journal} {\bibinfo  {journal} {Z. Phys. A}\ }\textbf {\bibinfo {volume}
  {339}},\ \bibinfo {pages} {23} (\bibinfo {year} {1991})}\BibitemShut
  {NoStop}%
\bibitem [{\citenamefont {Zhou}\ \emph {et~al.}(2003)\citenamefont {Zhou},
  \citenamefont {Meng},\ and\ \citenamefont {Ring}}]{Zhou2003PRC}%
  \BibitemOpen
  \bibfield  {author} {\bibinfo {author} {\bibfnamefont {S.-G.}\ \bibnamefont
  {Zhou}}, \bibinfo {author} {\bibfnamefont {J.}~\bibnamefont {Meng}}, \ and\
  \bibinfo {author} {\bibfnamefont {P.}~\bibnamefont {Ring}},\ }\href {\doibase
  10.1103/PhysRevC.68.034323} {\bibfield  {journal} {\bibinfo  {journal} {Phys.
  Rev. C}\ }\textbf {\bibinfo {volume} {68}},\ \bibinfo {pages} {034323}
  (\bibinfo {year} {2003})}\BibitemShut {NoStop}%
\bibitem [{\citenamefont {Zhang}\ \emph
  {et~al.}(2022{\natexlab{a}})\citenamefont {Zhang}, \citenamefont {Pan},\ and\
  \citenamefont {Zhang}}]{Zhang2022PRC}%
  \BibitemOpen
  \bibfield  {author} {\bibinfo {author} {\bibfnamefont {K.~Y.}\ \bibnamefont
  {Zhang}}, \bibinfo {author} {\bibfnamefont {C.}~\bibnamefont {Pan}}, \ and\
  \bibinfo {author} {\bibfnamefont {S.~Q.}\ \bibnamefont {Zhang}},\ }\href
  {\doibase 10.1103/PhysRevC.106.024302} {\bibfield  {journal} {\bibinfo
  {journal} {Phys. Rev. C}\ }\textbf {\bibinfo {volume} {106}},\ \bibinfo
  {pages} {024302} (\bibinfo {year} {2022}{\natexlab{a}})}\BibitemShut
  {NoStop}%
\bibitem [{\citenamefont {Ring}\ and\ \citenamefont
  {Schuck}(1980)}]{Peter1980Book}%
  \BibitemOpen
  \bibfield  {author} {\bibinfo {author} {\bibfnamefont {P.}~\bibnamefont
  {Ring}}\ and\ \bibinfo {author} {\bibfnamefont {P.}~\bibnamefont {Schuck}},\
  }\href@noop {} {\emph {\bibinfo {title} {The Nuclear Many-Body Problem}}}\
  (\bibinfo  {publisher} {Springer-Verlag, Berlin},\ \bibinfo {year}
  {1980})\BibitemShut {NoStop}%
\bibitem [{\citenamefont {Zhao}\ \emph {et~al.}(2010)\citenamefont {Zhao},
  \citenamefont {Li}, \citenamefont {Yao},\ and\ \citenamefont
  {Meng}}]{Zhao2010PRC}%
  \BibitemOpen
  \bibfield  {author} {\bibinfo {author} {\bibfnamefont {P.~W.}\ \bibnamefont
  {Zhao}}, \bibinfo {author} {\bibfnamefont {Z.~P.}\ \bibnamefont {Li}},
  \bibinfo {author} {\bibfnamefont {J.~M.}\ \bibnamefont {Yao}}, \ and\
  \bibinfo {author} {\bibfnamefont {J.}~\bibnamefont {Meng}},\ }\href {\doibase
  10.1103/PhysRevC.82.054319} {\bibfield  {journal} {\bibinfo  {journal} {Phys.
  Rev. C}\ }\textbf {\bibinfo {volume} {82}},\ \bibinfo {pages} {054319}
  (\bibinfo {year} {2010})}\BibitemShut {NoStop}%
\bibitem [{\citenamefont {Lalazissis}\ \emph {et~al.}(2009)\citenamefont
  {Lalazissis}, \citenamefont {Karatzikos}, \citenamefont {Fossion},
  \citenamefont {Arteaga}, \citenamefont {Afanasjev},\ and\ \citenamefont
  {Ring}}]{Lalazissis2009PLB}%
  \BibitemOpen
  \bibfield  {author} {\bibinfo {author} {\bibfnamefont {G.~A.}\ \bibnamefont
  {Lalazissis}}, \bibinfo {author} {\bibfnamefont {S.}~\bibnamefont
  {Karatzikos}}, \bibinfo {author} {\bibfnamefont {R.}~\bibnamefont {Fossion}},
  \bibinfo {author} {\bibfnamefont {D.~P.}\ \bibnamefont {Arteaga}}, \bibinfo
  {author} {\bibfnamefont {A.~V.}\ \bibnamefont {Afanasjev}}, \ and\ \bibinfo
  {author} {\bibfnamefont {P.}~\bibnamefont {Ring}},\ }\href {\doibase
  10.1016/j.physletb.2008.11.070} {\bibfield  {journal} {\bibinfo  {journal}
  {Phys. Lett. B}\ }\textbf {\bibinfo {volume} {671}},\ \bibinfo {pages} {36}
  (\bibinfo {year} {2009})}\BibitemShut {NoStop}%
\bibitem [{\citenamefont {Sharma}\ \emph {et~al.}(1993)\citenamefont {Sharma},
  \citenamefont {Nagarajan},\ and\ \citenamefont {Ring}}]{Sharma1993PLB}%
  \BibitemOpen
  \bibfield  {author} {\bibinfo {author} {\bibfnamefont {M.~M.}\ \bibnamefont
  {Sharma}}, \bibinfo {author} {\bibfnamefont {M.~A.}\ \bibnamefont
  {Nagarajan}}, \ and\ \bibinfo {author} {\bibfnamefont {P.}~\bibnamefont
  {Ring}},\ }\href {\doibase 10.1016/0370-2693(93)90970-S} {\bibfield
  {journal} {\bibinfo  {journal} {Phys. Lett. B}\ }\textbf {\bibinfo {volume}
  {312}},\ \bibinfo {pages} {377} (\bibinfo {year} {1993})}\BibitemShut
  {NoStop}%
\bibitem [{\citenamefont {Long}\ \emph {et~al.}(2004)\citenamefont {Long},
  \citenamefont {Meng}, \citenamefont {Van~Giai},\ and\ \citenamefont
  {Zhou}}]{Long2004PRC}%
  \BibitemOpen
  \bibfield  {author} {\bibinfo {author} {\bibfnamefont {W.}~\bibnamefont
  {Long}}, \bibinfo {author} {\bibfnamefont {J.}~\bibnamefont {Meng}}, \bibinfo
  {author} {\bibfnamefont {N.}~\bibnamefont {Van~Giai}}, \ and\ \bibinfo
  {author} {\bibfnamefont {S.-G.}\ \bibnamefont {Zhou}},\ }\href {\doibase
  10.1103/PhysRevC.69.034319} {\bibfield  {journal} {\bibinfo  {journal} {Phys.
  Rev. C}\ }\textbf {\bibinfo {volume} {69}},\ \bibinfo {pages} {034319}
  (\bibinfo {year} {2004})}\BibitemShut {NoStop}%
\bibitem [{\citenamefont {Xia}\ \emph {et~al.}(2018)\citenamefont {Xia},
  \citenamefont {Lim}, \citenamefont {Zhao}, \citenamefont {Liang},
  \citenamefont {Qu}, \citenamefont {Chen}, \citenamefont {Liu}, \citenamefont
  {Zhang}, \citenamefont {Zhang}, \citenamefont {Kim},\ and\ \citenamefont
  {Meng}}]{Xia2018ADNDT}%
  \BibitemOpen
  \bibfield  {author} {\bibinfo {author} {\bibfnamefont {X.~W.}\ \bibnamefont
  {Xia}}, \bibinfo {author} {\bibfnamefont {Y.}~\bibnamefont {Lim}}, \bibinfo
  {author} {\bibfnamefont {P.~W.}\ \bibnamefont {Zhao}}, \bibinfo {author}
  {\bibfnamefont {H.~Z.}\ \bibnamefont {Liang}}, \bibinfo {author}
  {\bibfnamefont {X.~Y.}\ \bibnamefont {Qu}}, \bibinfo {author} {\bibfnamefont
  {Y.}~\bibnamefont {Chen}}, \bibinfo {author} {\bibfnamefont {H.}~\bibnamefont
  {Liu}}, \bibinfo {author} {\bibfnamefont {L.~F.}\ \bibnamefont {Zhang}},
  \bibinfo {author} {\bibfnamefont {S.~Q.}\ \bibnamefont {Zhang}}, \bibinfo
  {author} {\bibfnamefont {Y.}~\bibnamefont {Kim}}, \ and\ \bibinfo {author}
  {\bibfnamefont {J.}~\bibnamefont {Meng}},\ }\href {\doibase
  10.1016/j.adt.2017.09.001} {\bibfield  {journal} {\bibinfo  {journal} {Atom.
  Data Nucl. Data Tabl.}\ }\textbf {\bibinfo {volume} {121-122}},\ \bibinfo
  {pages} {1} (\bibinfo {year} {2018})}\BibitemShut {NoStop}%
\bibitem [{\citenamefont {Zhang}\ \emph {et~al.}(2020)\citenamefont {Zhang},
  \citenamefont {Cheoun}, \citenamefont {Choi}, \citenamefont {Chong},
  \citenamefont {Dong}, \citenamefont {Geng}, \citenamefont {Ha}, \citenamefont
  {He}, \citenamefont {Heo}, \citenamefont {Ho}, \citenamefont {In},
  \citenamefont {Kim}, \citenamefont {Kim}, \citenamefont {Lee}, \citenamefont
  {Lee}, \citenamefont {Li}, \citenamefont {Luo}, \citenamefont {Meng},
  \citenamefont {Mun}, \citenamefont {Niu}, \citenamefont {Pan}, \citenamefont
  {Papakonstantinou}, \citenamefont {Shang}, \citenamefont {Shen},
  \citenamefont {Shen}, \citenamefont {Sun}, \citenamefont {Sun}, \citenamefont
  {Tam}, \citenamefont {Thaivayongnou}, \citenamefont {Wang}, \citenamefont
  {Wong}, \citenamefont {Xia}, \citenamefont {Yan}, \citenamefont {Yeung},
  \citenamefont {Yiu}, \citenamefont {Zhang}, \citenamefont {Zhang},\ and\
  \citenamefont {Zhou}}]{Zhang2020PRC}%
  \BibitemOpen
  \bibfield  {author} {\bibinfo {author} {\bibfnamefont {K.}~\bibnamefont
  {Zhang}}, \bibinfo {author} {\bibfnamefont {M.-K.}\ \bibnamefont {Cheoun}},
  \bibinfo {author} {\bibfnamefont {Y.-B.}\ \bibnamefont {Choi}}, \bibinfo
  {author} {\bibfnamefont {P.~S.}\ \bibnamefont {Chong}}, \bibinfo {author}
  {\bibfnamefont {J.}~\bibnamefont {Dong}}, \bibinfo {author} {\bibfnamefont
  {L.}~\bibnamefont {Geng}}, \bibinfo {author} {\bibfnamefont {E.}~\bibnamefont
  {Ha}}, \bibinfo {author} {\bibfnamefont {X.}~\bibnamefont {He}}, \bibinfo
  {author} {\bibfnamefont {C.}~\bibnamefont {Heo}}, \bibinfo {author}
  {\bibfnamefont {M.~C.}\ \bibnamefont {Ho}}, \bibinfo {author} {\bibfnamefont
  {E.~J.}\ \bibnamefont {In}}, \bibinfo {author} {\bibfnamefont
  {S.}~\bibnamefont {Kim}}, \bibinfo {author} {\bibfnamefont {Y.}~\bibnamefont
  {Kim}}, \bibinfo {author} {\bibfnamefont {C.-H.}\ \bibnamefont {Lee}},
  \bibinfo {author} {\bibfnamefont {J.}~\bibnamefont {Lee}}, \bibinfo {author}
  {\bibfnamefont {Z.}~\bibnamefont {Li}}, \bibinfo {author} {\bibfnamefont
  {T.}~\bibnamefont {Luo}}, \bibinfo {author} {\bibfnamefont {J.}~\bibnamefont
  {Meng}}, \bibinfo {author} {\bibfnamefont {M.-H.}\ \bibnamefont {Mun}},
  \bibinfo {author} {\bibfnamefont {Z.}~\bibnamefont {Niu}}, \bibinfo {author}
  {\bibfnamefont {C.}~\bibnamefont {Pan}}, \bibinfo {author} {\bibfnamefont
  {P.}~\bibnamefont {Papakonstantinou}}, \bibinfo {author} {\bibfnamefont
  {X.}~\bibnamefont {Shang}}, \bibinfo {author} {\bibfnamefont
  {C.}~\bibnamefont {Shen}}, \bibinfo {author} {\bibfnamefont {G.}~\bibnamefont
  {Shen}}, \bibinfo {author} {\bibfnamefont {W.}~\bibnamefont {Sun}}, \bibinfo
  {author} {\bibfnamefont {X.-X.}\ \bibnamefont {Sun}}, \bibinfo {author}
  {\bibfnamefont {C.~K.}\ \bibnamefont {Tam}}, \bibinfo {author} {\bibnamefont
  {Thaivayongnou}}, \bibinfo {author} {\bibfnamefont {C.}~\bibnamefont {Wang}},
  \bibinfo {author} {\bibfnamefont {S.~H.}\ \bibnamefont {Wong}}, \bibinfo
  {author} {\bibfnamefont {X.}~\bibnamefont {Xia}}, \bibinfo {author}
  {\bibfnamefont {Y.}~\bibnamefont {Yan}}, \bibinfo {author} {\bibfnamefont
  {R.~W.-Y.}\ \bibnamefont {Yeung}}, \bibinfo {author} {\bibfnamefont {T.~C.}\
  \bibnamefont {Yiu}}, \bibinfo {author} {\bibfnamefont {S.}~\bibnamefont
  {Zhang}}, \bibinfo {author} {\bibfnamefont {W.}~\bibnamefont {Zhang}}, \ and\
  \bibinfo {author} {\bibfnamefont {S.-G.}\ \bibnamefont {Zhou}} (\bibinfo
  {collaboration} {DRHBc Mass Table Collaboration}),\ }\href {\doibase
  10.1103/PhysRevC.102.024314} {\bibfield  {journal} {\bibinfo  {journal}
  {Phys. Rev. C}\ }\textbf {\bibinfo {volume} {102}},\ \bibinfo {pages}
  {024314} (\bibinfo {year} {2020})}\BibitemShut {NoStop}%
\bibitem [{\citenamefont {Pan}\ \emph {et~al.}(2019)\citenamefont {Pan},
  \citenamefont {Zhang},\ and\ \citenamefont {Zhang}}]{Pan2019IJMPE}%
  \BibitemOpen
  \bibfield  {author} {\bibinfo {author} {\bibfnamefont {C.}~\bibnamefont
  {Pan}}, \bibinfo {author} {\bibfnamefont {K.}~\bibnamefont {Zhang}}, \ and\
  \bibinfo {author} {\bibfnamefont {S.}~\bibnamefont {Zhang}},\ }\href
  {\doibase 10.1142/S0218301319500824} {\bibfield  {journal} {\bibinfo
  {journal} {Int. J. Mod. Phys. E}\ }\textbf {\bibinfo {volume} {28}},\
  \bibinfo {pages} {1950082} (\bibinfo {year} {2019})}\BibitemShut {NoStop}%
\bibitem [{\citenamefont {Zhang}\ \emph
  {et~al.}(2022{\natexlab{b}})\citenamefont {Zhang}, \citenamefont {Cheoun},
  \citenamefont {Choi}, \citenamefont {Chong}, \citenamefont {Dong},
  \citenamefont {Dong}, \citenamefont {Du}, \citenamefont {Geng}, \citenamefont
  {Ha}, \citenamefont {He}, \citenamefont {Heo}, \citenamefont {Ho},
  \citenamefont {In}, \citenamefont {Kim}, \citenamefont {Kim}, \citenamefont
  {Lee}, \citenamefont {Lee}, \citenamefont {Li}, \citenamefont {Li},
  \citenamefont {Luo}, \citenamefont {Meng}, \citenamefont {Mun}, \citenamefont
  {Niu}, \citenamefont {Pan}, \citenamefont {Papakonstantinou}, \citenamefont
  {Shang}, \citenamefont {Shen}, \citenamefont {Shen}, \citenamefont {Sun},
  \citenamefont {Sun}, \citenamefont {Tam}, \citenamefont {Thaivayongnou},
  \citenamefont {Wang}, \citenamefont {Wang}, \citenamefont {Wong},
  \citenamefont {Wu}, \citenamefont {Wu}, \citenamefont {Xia}, \citenamefont
  {Yan}, \citenamefont {Yeung}, \citenamefont {Yiu}, \citenamefont {Zhang},
  \citenamefont {Zhang}, \citenamefont {Zhang}, \citenamefont {Zhao},\ and\
  \citenamefont {Zhou}}]{Zhang2022ADNDT}%
  \BibitemOpen
  \bibfield  {author} {\bibinfo {author} {\bibfnamefont {K.}~\bibnamefont
  {Zhang}}, \bibinfo {author} {\bibfnamefont {M.-K.}\ \bibnamefont {Cheoun}},
  \bibinfo {author} {\bibfnamefont {Y.-B.}\ \bibnamefont {Choi}}, \bibinfo
  {author} {\bibfnamefont {P.~S.}\ \bibnamefont {Chong}}, \bibinfo {author}
  {\bibfnamefont {J.}~\bibnamefont {Dong}}, \bibinfo {author} {\bibfnamefont
  {Z.}~\bibnamefont {Dong}}, \bibinfo {author} {\bibfnamefont {X.}~\bibnamefont
  {Du}}, \bibinfo {author} {\bibfnamefont {L.}~\bibnamefont {Geng}}, \bibinfo
  {author} {\bibfnamefont {E.}~\bibnamefont {Ha}}, \bibinfo {author}
  {\bibfnamefont {X.-T.}\ \bibnamefont {He}}, \bibinfo {author} {\bibfnamefont
  {C.}~\bibnamefont {Heo}}, \bibinfo {author} {\bibfnamefont {M.~C.}\
  \bibnamefont {Ho}}, \bibinfo {author} {\bibfnamefont {E.~J.}\ \bibnamefont
  {In}}, \bibinfo {author} {\bibfnamefont {S.}~\bibnamefont {Kim}}, \bibinfo
  {author} {\bibfnamefont {Y.}~\bibnamefont {Kim}}, \bibinfo {author}
  {\bibfnamefont {C.-H.}\ \bibnamefont {Lee}}, \bibinfo {author} {\bibfnamefont
  {J.}~\bibnamefont {Lee}}, \bibinfo {author} {\bibfnamefont {H.}~\bibnamefont
  {Li}}, \bibinfo {author} {\bibfnamefont {Z.}~\bibnamefont {Li}}, \bibinfo
  {author} {\bibfnamefont {T.}~\bibnamefont {Luo}}, \bibinfo {author}
  {\bibfnamefont {J.}~\bibnamefont {Meng}}, \bibinfo {author} {\bibfnamefont
  {M.-H.}\ \bibnamefont {Mun}}, \bibinfo {author} {\bibfnamefont
  {Z.}~\bibnamefont {Niu}}, \bibinfo {author} {\bibfnamefont {C.}~\bibnamefont
  {Pan}}, \bibinfo {author} {\bibfnamefont {P.}~\bibnamefont
  {Papakonstantinou}}, \bibinfo {author} {\bibfnamefont {X.}~\bibnamefont
  {Shang}}, \bibinfo {author} {\bibfnamefont {C.}~\bibnamefont {Shen}},
  \bibinfo {author} {\bibfnamefont {G.}~\bibnamefont {Shen}}, \bibinfo {author}
  {\bibfnamefont {W.}~\bibnamefont {Sun}}, \bibinfo {author} {\bibfnamefont
  {X.-X.}\ \bibnamefont {Sun}}, \bibinfo {author} {\bibfnamefont {C.~K.}\
  \bibnamefont {Tam}}, \bibinfo {author} {\bibnamefont {Thaivayongnou}},
  \bibinfo {author} {\bibfnamefont {C.}~\bibnamefont {Wang}}, \bibinfo {author}
  {\bibfnamefont {X.}~\bibnamefont {Wang}}, \bibinfo {author} {\bibfnamefont
  {S.~H.}\ \bibnamefont {Wong}}, \bibinfo {author} {\bibfnamefont
  {J.}~\bibnamefont {Wu}}, \bibinfo {author} {\bibfnamefont {X.}~\bibnamefont
  {Wu}}, \bibinfo {author} {\bibfnamefont {X.}~\bibnamefont {Xia}}, \bibinfo
  {author} {\bibfnamefont {Y.}~\bibnamefont {Yan}}, \bibinfo {author}
  {\bibfnamefont {R.~W.-Y.}\ \bibnamefont {Yeung}}, \bibinfo {author}
  {\bibfnamefont {T.~C.}\ \bibnamefont {Yiu}}, \bibinfo {author} {\bibfnamefont
  {S.}~\bibnamefont {Zhang}}, \bibinfo {author} {\bibfnamefont
  {W.}~\bibnamefont {Zhang}}, \bibinfo {author} {\bibfnamefont
  {X.}~\bibnamefont {Zhang}}, \bibinfo {author} {\bibfnamefont
  {Q.}~\bibnamefont {Zhao}}, \ and\ \bibinfo {author} {\bibfnamefont {S.-G.}\
  \bibnamefont {Zhou}} (\bibinfo {collaboration} {DRHBc Mass Table
  Collaboration}),\ }\href {\doibase https://doi.org/10.1016/j.adt.2022.101488}
  {\bibfield  {journal} {\bibinfo  {journal} {Atom. Data Nucl. Data Tabl.}\
  }\textbf {\bibinfo {volume} {144}},\ \bibinfo {pages} {101488} (\bibinfo
  {year} {2022}{\natexlab{b}})}\BibitemShut {NoStop}%
\bibitem [{\citenamefont {Perez-Martin}\ and\ \citenamefont
  {Robledo}(2008)}]{Perez-Martin2008PRC}%
  \BibitemOpen
  \bibfield  {author} {\bibinfo {author} {\bibfnamefont {S.}~\bibnamefont
  {Perez-Martin}}\ and\ \bibinfo {author} {\bibfnamefont {L.~M.}\ \bibnamefont
  {Robledo}},\ }\href {\doibase 10.1103/PhysRevC.78.014304} {\bibfield
  {journal} {\bibinfo  {journal} {Phys. Rev. C}\ }\textbf {\bibinfo {volume}
  {78}},\ \bibinfo {pages} {014304} (\bibinfo {year} {2008})}\BibitemShut
  {NoStop}%
\bibitem [{\citenamefont {Li}\ \emph {et~al.}(2012{\natexlab{b}})\citenamefont
  {Li}, \citenamefont {Meng}, \citenamefont {Ring}, \citenamefont {Zhao},\ and\
  \citenamefont {Zhou}}]{Li2012CPL}%
  \BibitemOpen
  \bibfield  {author} {\bibinfo {author} {\bibfnamefont {L.}~\bibnamefont
  {Li}}, \bibinfo {author} {\bibfnamefont {J.}~\bibnamefont {Meng}}, \bibinfo
  {author} {\bibfnamefont {P.}~\bibnamefont {Ring}}, \bibinfo {author}
  {\bibfnamefont {E.-G.}\ \bibnamefont {Zhao}}, \ and\ \bibinfo {author}
  {\bibfnamefont {S.-G.}\ \bibnamefont {Zhou}},\ }\href {\doibase
  10.1088/0256-307X/29/4/042101} {\bibfield  {journal} {\bibinfo  {journal}
  {Chin. Phys. Lett.}\ }\textbf {\bibinfo {volume} {29}},\ \bibinfo {pages}
  {042101} (\bibinfo {year} {2012}{\natexlab{b}})}\BibitemShut {NoStop}%
\bibitem [{\citenamefont {Pan}\ \emph {et~al.}(2022)\citenamefont {Pan},
  \citenamefont {Cheoun}, \citenamefont {Choi}, \citenamefont {Dong},
  \citenamefont {Du}, \citenamefont {Fan}, \citenamefont {Gao}, \citenamefont
  {Geng}, \citenamefont {Ha}, \citenamefont {He}, \citenamefont {Huang},
  \citenamefont {Huang}, \citenamefont {Kim}, \citenamefont {Kim},
  \citenamefont {Lee}, \citenamefont {Lee}, \citenamefont {Li}, \citenamefont
  {Liu}, \citenamefont {Ma}, \citenamefont {Meng}, \citenamefont {Mun},
  \citenamefont {Niu}, \citenamefont {Papakonstantinou}, \citenamefont {Shang},
  \citenamefont {Shen}, \citenamefont {Shen}, \citenamefont {Sun},
  \citenamefont {Sun}, \citenamefont {Wu}, \citenamefont {Wu}, \citenamefont
  {Xia}, \citenamefont {Yan}, \citenamefont {Yiu}, \citenamefont {Zhang},
  \citenamefont {Zhang}, \citenamefont {Zhang}, \citenamefont {Zhang},
  \citenamefont {Zhao}, \citenamefont {Zheng},\ and\ \citenamefont
  {Zhou}}]{Pan2022PRC}%
  \BibitemOpen
  \bibfield  {author} {\bibinfo {author} {\bibfnamefont {C.}~\bibnamefont
  {Pan}}, \bibinfo {author} {\bibfnamefont {M.-K.}\ \bibnamefont {Cheoun}},
  \bibinfo {author} {\bibfnamefont {Y.-B.}\ \bibnamefont {Choi}}, \bibinfo
  {author} {\bibfnamefont {J.}~\bibnamefont {Dong}}, \bibinfo {author}
  {\bibfnamefont {X.}~\bibnamefont {Du}}, \bibinfo {author} {\bibfnamefont
  {X.-H.}\ \bibnamefont {Fan}}, \bibinfo {author} {\bibfnamefont
  {W.}~\bibnamefont {Gao}}, \bibinfo {author} {\bibfnamefont {L.}~\bibnamefont
  {Geng}}, \bibinfo {author} {\bibfnamefont {E.}~\bibnamefont {Ha}}, \bibinfo
  {author} {\bibfnamefont {X.-T.}\ \bibnamefont {He}}, \bibinfo {author}
  {\bibfnamefont {J.}~\bibnamefont {Huang}}, \bibinfo {author} {\bibfnamefont
  {K.}~\bibnamefont {Huang}}, \bibinfo {author} {\bibfnamefont
  {S.}~\bibnamefont {Kim}}, \bibinfo {author} {\bibfnamefont {Y.}~\bibnamefont
  {Kim}}, \bibinfo {author} {\bibfnamefont {C.-H.}\ \bibnamefont {Lee}},
  \bibinfo {author} {\bibfnamefont {J.}~\bibnamefont {Lee}}, \bibinfo {author}
  {\bibfnamefont {Z.}~\bibnamefont {Li}}, \bibinfo {author} {\bibfnamefont
  {Z.-R.}\ \bibnamefont {Liu}}, \bibinfo {author} {\bibfnamefont
  {Y.}~\bibnamefont {Ma}}, \bibinfo {author} {\bibfnamefont {J.}~\bibnamefont
  {Meng}}, \bibinfo {author} {\bibfnamefont {M.-H.}\ \bibnamefont {Mun}},
  \bibinfo {author} {\bibfnamefont {Z.}~\bibnamefont {Niu}}, \bibinfo {author}
  {\bibfnamefont {P.}~\bibnamefont {Papakonstantinou}}, \bibinfo {author}
  {\bibfnamefont {X.}~\bibnamefont {Shang}}, \bibinfo {author} {\bibfnamefont
  {C.}~\bibnamefont {Shen}}, \bibinfo {author} {\bibfnamefont {G.}~\bibnamefont
  {Shen}}, \bibinfo {author} {\bibfnamefont {W.}~\bibnamefont {Sun}}, \bibinfo
  {author} {\bibfnamefont {X.-X.}\ \bibnamefont {Sun}}, \bibinfo {author}
  {\bibfnamefont {J.}~\bibnamefont {Wu}}, \bibinfo {author} {\bibfnamefont
  {X.}~\bibnamefont {Wu}}, \bibinfo {author} {\bibfnamefont {X.}~\bibnamefont
  {Xia}}, \bibinfo {author} {\bibfnamefont {Y.}~\bibnamefont {Yan}}, \bibinfo
  {author} {\bibfnamefont {T.~C.}\ \bibnamefont {Yiu}}, \bibinfo {author}
  {\bibfnamefont {K.}~\bibnamefont {Zhang}}, \bibinfo {author} {\bibfnamefont
  {S.}~\bibnamefont {Zhang}}, \bibinfo {author} {\bibfnamefont
  {W.}~\bibnamefont {Zhang}}, \bibinfo {author} {\bibfnamefont
  {X.}~\bibnamefont {Zhang}}, \bibinfo {author} {\bibfnamefont
  {Q.}~\bibnamefont {Zhao}}, \bibinfo {author} {\bibfnamefont {R.}~\bibnamefont
  {Zheng}}, \ and\ \bibinfo {author} {\bibfnamefont {S.-G.}\ \bibnamefont
  {Zhou}} (\bibinfo {collaboration} {DRHBc Mass Table Collaboration}),\ }\href
  {\doibase 10.1103/PhysRevC.106.014316} {\bibfield  {journal} {\bibinfo
  {journal} {Phys. Rev. C}\ }\textbf {\bibinfo {volume} {106}},\ \bibinfo
  {pages} {014316} (\bibinfo {year} {2022})}\BibitemShut {NoStop}%
\bibitem [{\citenamefont {Wang}\ \emph {et~al.}(2021)\citenamefont {Wang},
  \citenamefont {Huang}, \citenamefont {Kondev}, \citenamefont {Audi},\ and\
  \citenamefont {Naimi}}]{AME2020(3)}%
  \BibitemOpen
  \bibfield  {author} {\bibinfo {author} {\bibfnamefont {M.}~\bibnamefont
  {Wang}}, \bibinfo {author} {\bibfnamefont {W.}~\bibnamefont {Huang}},
  \bibinfo {author} {\bibfnamefont {F.}~\bibnamefont {Kondev}}, \bibinfo
  {author} {\bibfnamefont {G.}~\bibnamefont {Audi}}, \ and\ \bibinfo {author}
  {\bibfnamefont {S.}~\bibnamefont {Naimi}},\ }\href
  {https://iopscience.iop.org/article/10.1088/1674-1137/abddaf/meta} {\bibfield
   {journal} {\bibinfo  {journal} {Chin. Phys. C}\ }\textbf {\bibinfo {volume}
  {45}},\ \bibinfo {pages} {030003} (\bibinfo {year} {2021})}\BibitemShut
  {NoStop}%
\bibitem [{\citenamefont {Heylen}\ \emph {et~al.}(2021)\citenamefont {Heylen},
  \citenamefont {Devlin}, \citenamefont {Gins}, \citenamefont {Bissell},
  \citenamefont {Blaum}, \citenamefont {Cheal}, \citenamefont {Filippin},
  \citenamefont {Ruiz}, \citenamefont {Godefroid}, \citenamefont {Gorges},
  \citenamefont {Holt}, \citenamefont {Kanellakopoulos}, \citenamefont
  {Kaufmann}, \citenamefont {Koszor\'us}, \citenamefont {K\"onig},
  \citenamefont {Malbrunot-Ettenauer}, \citenamefont {Miyagi}, \citenamefont
  {Neugart}, \citenamefont {Neyens}, \citenamefont {N\"ortersh\"auser},
  \citenamefont {S\'anchez}, \citenamefont {Sommer}, \citenamefont
  {Rodr\'{\i}guez}, \citenamefont {Xie}, \citenamefont {Xu}, \citenamefont
  {Yang},\ and\ \citenamefont {Yordanov}}]{Heylen2021PRC}%
  \BibitemOpen
  \bibfield  {author} {\bibinfo {author} {\bibfnamefont {H.}~\bibnamefont
  {Heylen}}, \bibinfo {author} {\bibfnamefont {C.~S.}\ \bibnamefont {Devlin}},
  \bibinfo {author} {\bibfnamefont {W.}~\bibnamefont {Gins}}, \bibinfo {author}
  {\bibfnamefont {M.~L.}\ \bibnamefont {Bissell}}, \bibinfo {author}
  {\bibfnamefont {K.}~\bibnamefont {Blaum}}, \bibinfo {author} {\bibfnamefont
  {B.}~\bibnamefont {Cheal}}, \bibinfo {author} {\bibfnamefont
  {L.}~\bibnamefont {Filippin}}, \bibinfo {author} {\bibfnamefont {R.~F.~Garcia}\
  \bibnamefont {Ruiz}}, \bibinfo {author} {\bibfnamefont {M.}~\bibnamefont
  {Godefroid}}, \bibinfo {author} {\bibfnamefont {C.}~\bibnamefont {Gorges}},
  \bibinfo {author} {\bibfnamefont {J.~D.}\ \bibnamefont {Holt}}, \bibinfo
  {author} {\bibfnamefont {A.}~\bibnamefont {Kanellakopoulos}}, \bibinfo
  {author} {\bibfnamefont {S.}~\bibnamefont {Kaufmann}}, \bibinfo {author}
  {\bibfnamefont {A.}~\bibnamefont {Koszor\'us}}, \bibinfo {author}
  {\bibfnamefont {K.}~\bibnamefont {K\"onig}}, \bibinfo {author} {\bibfnamefont
  {S.}~\bibnamefont {Malbrunot-Ettenauer}}, \bibinfo {author} {\bibfnamefont
  {T.}~\bibnamefont {Miyagi}}, \bibinfo {author} {\bibfnamefont
  {R.}~\bibnamefont {Neugart}}, \bibinfo {author} {\bibfnamefont
  {G.}~\bibnamefont {Neyens}}, \bibinfo {author} {\bibfnamefont
  {W.}~\bibnamefont {N\"ortersh\"auser}}, \bibinfo {author} {\bibfnamefont
  {R.}~\bibnamefont {S\'anchez}}, \bibinfo {author} {\bibfnamefont
  {F.}~\bibnamefont {Sommer}}, \bibinfo {author} {\bibfnamefont {L.~V.}\
  \bibnamefont {Rodr\'{\i}guez}}, \bibinfo {author} {\bibfnamefont
  {L.}~\bibnamefont {Xie}}, \bibinfo {author} {\bibfnamefont {Z.~Y.}\
  \bibnamefont {Xu}}, \bibinfo {author} {\bibfnamefont {X.~F.}\ \bibnamefont
  {Yang}}, \ and\ \bibinfo {author} {\bibfnamefont {D.~T.}\ \bibnamefont
  {Yordanov}},\ }\href {\doibase 10.1103/PhysRevC.103.014318} {\bibfield
  {journal} {\bibinfo  {journal} {Phys. Rev. C}\ }\textbf {\bibinfo {volume}
  {103}},\ \bibinfo {pages} {014318} (\bibinfo {year} {2021})}\BibitemShut
  {NoStop}%
\bibitem [{\citenamefont {Reinhard}(1988)}]{Reinhard1988ZPA}%
  \BibitemOpen
  \bibfield  {author} {\bibinfo {author} {\bibfnamefont {P.~G.}\ \bibnamefont
  {Reinhard}},\ }\href {\doibase 10.1007/BF01290231} {\bibfield  {journal}
  {\bibinfo  {journal} {Z. Phys. A}\ }\textbf {\bibinfo {volume} {329}},\
  \bibinfo {pages} {257} (\bibinfo {year} {1988})}\BibitemShut {NoStop}%
\bibitem [{\citenamefont {Baumann}\ \emph {et~al.}(2007)\citenamefont
  {Baumann}, \citenamefont {Amthor}, \citenamefont {Bazin}, \citenamefont
  {Brown}, \citenamefont {C.~M.~Folden}, \citenamefont {Gade}, \citenamefont
  {Ginter}, \citenamefont {Hausmann}, \citenamefont {Matos}, \citenamefont
  {Morrissey}, \citenamefont {Portillo}, \citenamefont {Schiller},
  \citenamefont {Sherrill}, \citenamefont {Stolz}, \citenamefont {Tarasov},\
  and\ \citenamefont {Thoennessen}}]{Baumann2007Nature}%
  \BibitemOpen
  \bibfield  {author} {\bibinfo {author} {\bibfnamefont {T.}~\bibnamefont
  {Baumann}}, \bibinfo {author} {\bibfnamefont {A.~M.}\ \bibnamefont {Amthor}},
  \bibinfo {author} {\bibfnamefont {D.}~\bibnamefont {Bazin}}, \bibinfo
  {author} {\bibfnamefont {B.~A.}\ \bibnamefont {Brown}}, \bibinfo {author}
  {\bibfnamefont {I.~I.~I.}\ \bibnamefont {C.~M.~Folden}}, \bibinfo {author}
  {\bibfnamefont {A.}~\bibnamefont {Gade}}, \bibinfo {author} {\bibfnamefont
  {T.~N.}\ \bibnamefont {Ginter}}, \bibinfo {author} {\bibfnamefont
  {M.}~\bibnamefont {Hausmann}}, \bibinfo {author} {\bibfnamefont
  {M.}~\bibnamefont {Matos}}, \bibinfo {author} {\bibfnamefont {D.~J.}\
  \bibnamefont {Morrissey}}, \bibinfo {author} {\bibfnamefont {M.}~\bibnamefont
  {Portillo}}, \bibinfo {author} {\bibfnamefont {A.}~\bibnamefont {Schiller}},
  \bibinfo {author} {\bibfnamefont {B.~M.}\ \bibnamefont {Sherrill}}, \bibinfo
  {author} {\bibfnamefont {A.}~\bibnamefont {Stolz}}, \bibinfo {author}
  {\bibfnamefont {O.~B.}\ \bibnamefont {Tarasov}}, \ and\ \bibinfo {author}
  {\bibfnamefont {M.}~\bibnamefont {Thoennessen}},\ }\href {\doibase
  10.1038/nature06213} {\bibfield  {journal} {\bibinfo  {journal} {Nature}\
  }\textbf {\bibinfo {volume} {449}},\ \bibinfo {pages} {1022} (\bibinfo {year}
  {2007})}\BibitemShut {NoStop}%
\bibitem [{\citenamefont {Tanihata}\ \emph {et~al.}(2013)\citenamefont
  {Tanihata}, \citenamefont {Savajols},\ and\ \citenamefont
  {Kanungo}}]{Tanihata2013PPNP}%
  \BibitemOpen
  \bibfield  {author} {\bibinfo {author} {\bibfnamefont {I.}~\bibnamefont
  {Tanihata}}, \bibinfo {author} {\bibfnamefont {H.}~\bibnamefont {Savajols}},
  \ and\ \bibinfo {author} {\bibfnamefont {R.}~\bibnamefont {Kanungo}},\ }\href
  {\doibase https://doi.org/10.1016/j.ppnp.2012.07.001} {\bibfield  {journal}
  {\bibinfo  {journal} {Prog. Part. Nucl. Phys.}\ }\textbf {\bibinfo {volume}
  {68}},\ \bibinfo {pages} {215 } (\bibinfo {year} {2013})}\BibitemShut
  {NoStop}%
\bibitem [{\citenamefont {Meng}\ and\ \citenamefont
  {Zhou}(2015)}]{Meng2015JPG}%
  \BibitemOpen
  \bibfield  {author} {\bibinfo {author} {\bibfnamefont {J.}~\bibnamefont
  {Meng}}\ and\ \bibinfo {author} {\bibfnamefont {S.~G.}\ \bibnamefont
  {Zhou}},\ }\href {\doibase 10.1088/0954-3899/42/9/093101} {\bibfield
  {journal} {\bibinfo  {journal} {J. Phys. G}\ }\textbf {\bibinfo {volume}
  {42}},\ \bibinfo {pages} {093101} (\bibinfo {year} {2015})}\BibitemShut
  {NoStop}%
\end{thebibliography}
%

\end{CJK*}
\end{document}